\documentclass[preprint]{aastex}

\shorttitle{Multiple Super-Earth formation}

\shortauthors{Ida and Lin}

\begin{document}

\title{Toward a Deterministic Model of Planetary Formation VI:}
\title{Dynamical Interaction and Coagulation of Multiple Rocky 
Embryos and Super-Earth Systems around Solar Type Stars}

\author{S. Ida}
\affil{Tokyo Institute of Technology,
Ookayama, Meguro-ku, Tokyo 152-8551, Japan}
\email{ida@geo.titech.ac.jp}

\and 

\author{D. N. C. Lin}
\affil{UCO/Lick Observatory, University of California, 
Santa Cruz, CA 95064}
\affil{Kavli Institute for Astronomy and Astrophysics, Peking 
University, Beijing, China}
\email{lin@ucolick.org}

\begin{abstract}
Radial velocity and transit surveys indicate that solar-type stars
bear super-Earths, with mass and period up to $\sim 20 M_\oplus$ and a
few months, are more common than those with Jupiter-mass gas giants.
In many cases, these super-Earths are members of multiple-planet
systems in which their mutual dynamical interaction has influenced their 
formation and evolution. In this paper, we modify an existing numerical
population synthesis scheme to take into account protoplanetary
embryos' interaction with their evolving natal gaseous disk, as well as their
close scatterings and resonant interaction with each other. We show
that it is possible for a group of compact embryos to emerge interior
to the ice line, grow, migrate, and congregate into closely-packed
convoys which stall in the proximity of their host stars.  After the
disk-gas depletion, they undergo orbit crossing, close scattering, and
giant impacts to form multiple rocky Earths or super-Earths in
non-resonant orbits around $\sim 0.1$AU with moderate eccentricities
of $\sim 0.01$--0.1. We suggest that most refractory super-Earths 
with period in the range of a few days to weeks may have formed through 
this process.  These super-Earths differ from Neptune-like ice giants 
by their compact sizes and lack of a substantial gaseous envelope.
\end{abstract}

\section{Introduction}
In the past decade, more than 400 extrasolar planets have been
discovered around nearby solar-type stars.  Most of these known 
planets are probably gas giants because they have masses ($M_p$) 
and densities ($\rho_p$) comparable to those of Jupiter and Saturn, 
while their periods ($P$) ranges from a few days to years.  
When this population is extrapolated to longer periods (more than 
a decades), their ubiquity implies that the fraction of solar-type 
stars which have gas giant planets around them is approximately 
$\eta_J \sim 10$--$15\%$\citep{Cumming08}.  

Gas giant planets must formed in gas-rich environment.  However, 
in contrast to the modest value of $\eta_J$, the T Tauri progenitors 
of solar-type stars are commonly surrounded by protostellar disks 
\citep{Beckwith96} with surface densities comparable to that of 
the minimum mass solar nebula model (MMSN) \citep{Hayashi81}. Based on 
the conventional paradigm that gas giant planets form in frequently-observed,
evolving, natal protostellar disks through the condensation of grains,
coagulation of planetesimals, mergers of protoplanetary embryos, and gas
accretion onto super-Earth (with masses in excess of a few Earth-masses)
cores, we constructed a population-synthesis model through a series of 
papers \citep[][, hereafter Papers I-V]{IL04a,IL04b,IL05,IL08a,IL08b}.  In
these papers, we qualitatively showed that the low fertility rate of gas 
giants ({\it i.e.,} the modest value of $\eta_J$) around solar type stars despite the 
omnipresence of planetary cradles around their progenitor T Tauri stars 
is due to 1) an inefficient retention of building-block materials in 
protostellar disks and 2) that gas giant planet formation is only 
possible in relatively-massive (super-MMSN) protostellar disks which 
deplete on time scales ($\tau_{\rm dep}$) longer than $\sim 3$ Myr.

The observed mass-period ($M_p-P$) distribution of these planets appear 
to be non-uniform \citep{Udry_Santos, Cumming08}.  With 
a set of well-define and carefully-selected sample, potential observational 
selection effects may be quantitatively identified and taken into account
\citep{Schlaufman09}. Analogous to the stellar color-magnitude diagram for 
galactic and globular clusters, genuine oasis and deserts in planets' 
$M_p-P$ distribution of a controlled samples can be used to cast constraints 
and provide clues to the theory of planet formation and planetary-system 
evolution. 

Preliminary results of our first sets of simulations essentially reproduced 
known-planets' observed $M_p-P$ distribution and the fraction of 
planet-bearing stars as a function of their mass and metallicity (Papers II 
and III).  They have also been reproduced and confirmed by similar 
population-synthesis approaches \citep{Mordasini09} and hybrid (N-body 
$+$ 1D viscous-disk evolution) simulations\citep{Thommes08}, albeit there 
remain some differences in the asymptotic $M_p-P$ distribution. Nevertheless 
the statistical significance of these agreements remain unsettled, 
because 1) existing observational data are acquired with heterogeneous 
precession and diverse observational selection criteria and 2) theoretical
frameworks of some critical physical processes remain poorly understood.  
Ideally, the construction and upgrades of population-synthesis models 
can be used 1) to set quantitative calibrations on important model parameters
and 2) to extrapolate  tests and predictions for subsequent observations.

In our previous five papers, quantitative determination of the
efficiency of gas giant formation around different types of host stars, 
many uncertainties remain in the prescriptions adopted for these models.  
They include the magnitude of the following input parameters as defined 
in paper IV:

\noindent
1) the surface density distribution of gas (the fiducial magnitude
$\Sigma_g$ or $f_{g,0}$ and power index $q_g$ as defined in
Eq.~[\ref{eq:sigma_gas}] below), effective viscosity ($\nu$ or the
dimensionless $\alpha$ parameter) as a function of radius $r$ (see
Paper V), mass-flow rate ($\dot M_{\rm disk}$) and depletion time
scale $\tau_{\rm dep}$ of the disk gas (as defined Paper I),

\noindent
2) the surface density distribution of solid building-block (grains,
planetesimals, and protoplanetary embryos) material (the fiducial
magnitude $\Sigma_d$ or $f_d$ and power index $q_d$ as defined in
Eq.~[\ref{eq:sigma_dust}]),

\noindent
3) the scaling coefficients ($h_d$ and $h_g$ as defined in Papers II
and III) of dust and gas contents in the disk associated with a range
of the stellar mass ($M_\ast$) and metallicity ([Fe/H]),

\noindent
4) the rate of type-I migration of protoplanetary embryos which is
characterized by a parameter $C_1$ in comparison with the linear
calculation for unperturbed disks (see Paper IV), 

\noindent
5) the stalling process and location of both type I and II migration,
and

\noindent
6) the rate of gas accretion onto protoplanetary cores (which is
characterized by $k_1$ and $k_2$ in Paper I).

These properties are determined by the not-so-well understood
physical processes such as the efficiency of turbulent angular momentum 
transport (items 1 \& 3) and retention of heavy elements in the form of 
grains (items 2 \& 3), planetesimals, and embryos (items 4 \& 5). 
With regard to item 6, in the sequential accretion hypothesis, the 
formation of gas giants must be preceded by the emergence of cores 
with several $M_\oplus$'s \citep{Mizuno80, Pollack96, Ikoma00}. 
These uncertainties are not independent of each other. In particular,
the fraction of solar-type stars which bear gas giants depends on 
the inventory of building block material and the retention efficiency 
of cores.  In Paper IV, we demonstrated that in disks with power-law 
$\Sigma_g$ and $\Sigma_d$ distributions, the observed $\eta_J$ would 
be achievable only if 1) the disk is more massive than the MMSN model 
({\it i.e.} $f_d \ga 1$), and 2) type I migration is relatively 
inefficient ({\it i.e.} $C_1 \la 0.03$).  

These requirements need not be satisfied
throughout the disk provided there are special locations in the disk
where building-block grains can accumulate and embryos' type I
migration may be suppressed (see Paper V).  In this context, the ice
line appears to be a preferred birth place for gas giants because it
provides a natural barrier against the orbital decay of grains (due to
hydro dynamic drag) and protoplanetary embryos (due to type I
migration) \citep{Kretke07}.  The period distribution of the
simulated models which take into account of the preferred birth place
of gas giants (Paper V) can match well with a conspicuous (2-3 yrs)
up-turn in observed period distributions of Jupiter-mass planets
provided $C_1 \sim 0.1$ \citep{Schlaufman09}.

Ongoing transit searches and high-precision radial velocity surveys have
led to the discovery of many short-period (with $P$ less than a week) gas
giants. According to the population synthesis models, these planets
relocated to their present-day orbit either through type II migration or
dynamical instability and relaxation.  For this scenario, we need to 
verify: 1) criteria for gas giants' relocation shortly after
their formation, 2) efficiency of relocation processes, 3) condition
for stalling dynamical evolution, and 4) extent of gas giants' mass loss
and orbital decay during the main sequence life span of their host stars.
Through detailed quantitative comparisons between population-synthesis 
models and observational data, we can calibrate the efficiency of various
relevant and competing processes.

In addition to the reproduction of known observational properties (mostly
for gas giants), population-synthesis models may also be used to make 
some robust predictions. A particularly noticeable feature in the simulated
$M_p-P$ distribution is a domain of planetary desert for intermediate
mass ($\sim 20-100 M_\oplus$) and period (weeks to months) planets
with comparable gas and solid internal composition.
While the actual boundaries of this parameter region depend
on the values of various input parameters (such as $k_1$, $k_2$, $C_1$
etc), the paucity of planets in this domain is robust, mainly due to
the combined effects of runaway gas accretion (Paper I) and type I 
migration (Paper IV).  This region of parameter space would be filled 
if the gas accretion onto the cores is significantly slowed down by 
inefficient heat transfer in the infalling envelope or type I migration 
is grossly inhibited everywhere in the case where planetesimal accretion 
of migrating cores is not inhibited\citep{Mordasini09}. Thus, 
forthcoming observations can be used to distinguish key assumptions 
associated with these models.

In the context of quantitative predictions, the simulated planet 
distribution also highlights rich populations of both Earth-mass rocky 
planets and long-period ice giants. During the early epochs of disks' 
evolution (especially when they undergo repeated FU Ori outbursts), 
$\dot M_{\rm disk}$, $\Sigma_g$ and $\Sigma_d$ are much
larger than their values in the MMSN model. In principle, embryos can
rapidly undergo oligarchic growth \citep{KI98,KI02} and attain
isolation masses $M_{\rm c,iso}$ which are sufficiently large ($\sim
3-10 M_\oplus$) to initiate dynamical gas accretion exterior
to the ice line (Paper I). However, even with a modest amount of type I
migration ({\it i.e.} $C_1 \sim 10^{-2}$), the embryos would migrate
towards their host stars before they reach their isolation masses
(Paper IV). In inner regions at $\la 0.5$AU, the surface density of 
rocky materials is severely depleted by type I migration of embryos 
that start migration after they acquire mass comparable to their 
isolation mass (Paper IV).  

The fate of these embryos is determined by their retention probability.
If they are efficiently retained, they
would emerge as conspicuous super-Earths.  In contrast to gas giants'
type II migration, embryos' type I migration and retention processes
in the proximity of their host stars are poorly understood.  In Papers
I-V, we counted the number of embryos (with different masses) which
are expected to cross some inhibited inner disk boundary. 

In this paper, we consider various physical processes which may regulate
embryos' dynamical evolution and determine the formation and retention
probability of super-Earths at various locations.  This follow-up study 
is timely because recent radial-velocity surveys with high-cadence 
(several time nightly) and precision (down to $\sim$ m/s) have led to 
the announced discoveries of several super-Earths with masses and 
periods up to $\sim 20 M_\oplus$ and a few months (Mayor {\it et al.} 
2009). These data also suggest super-Earths may exist around nearly 
half of all nearby solar-type stars and in many cases in the form of 
non-resonant multiple-planet systems
(M. Mayor, S. Udry, private communication).
This trend, when confirmed with 
a well-defined controlled sample, would suggest there may be many 
Earth-mass or sub-Earth slowly-growing cores which failed to accrete 
any significant amount of gas before their natal disk is evaporated.
They also highlight the importance of dynamical interaction between 
multiple embryos during their formation and subsequent evolution.

In order to analyze the asymptotic evolution of multiple-embryo systems, 
we briefly summarize, in \S2, some relevant physical processes and 
introduce a set of generic prescription which can be reliably 
used to quantitative approximate embryos' interaction with each
other and their natal disks.  In \S3, we utilize this prescription, 
along with our previously
established algorithms, to carry out some case studies for formation
of multiple rocky/icy planetary systems around a majority of solar
type stars which do not bear any gas giant planets.  Based on the
results of Paper I and IV, we suggest these planets formed in
relatively low or modest mass disks. These disks generally have gas
surface density ($\Sigma_g$) less than 2-3 times that of the minimum
mass solar nebula (MMSN) model. We show that prior to gas depletion,
dynamically isolated embryos emerge with sub-Earth masses and nearly
circular orbits in the inner parts of these disks ({\it i.e.}
interior to the ice line). We refer to this first epoch as the
embryos-emergence stage.

During the (second) migration and accumulation stage, these embryos
rapidly migrate to proximity of their host stars. Under various
circumstances, embryos' migration may be stalled in the inner regions
of disks around classical T Tauri stars.  After the gas depletion, the
congregated embryos perturb each other's orbits.  Within the next
$\sim 100$ Myr, dynamical instability induces embryos to cross each
other's orbits and growth through cohesive collisions.  During this
(third) giant impact stage, super-Earth form with periods between a
few days to months and modest eccentricities ($e \sim 0.01$--0.1). As the
host stars mature, inelastic and cohesive collisions between embryos
with modest relative periapse longitudes damp their eccentricities.
In the absence of gas giants, systems eventually evolve into a
(fourth) stabilizing state in which the residual planets' crossing
timescale becomes larger than the age of the system.  We show that
these models naturally accounts for the origin of a known population
of non-resonant multiple Earth/super-Earth systems with periods ranging 
from days to months. The asymptotic mass of some planets may exceed
that (several Earth masses) needed to initiate efficient gas accretion
in a MMSN environment.  But these planets acquired most of their masses
through giant impacts after the gas is depleted from their natal disks.
In the absence of gas accretion, such super-Earths do not evolve into
gas giants.  Finally, we summarize our results and discuss their 
implications in \S4.

\section{Embryos interaction with each other and their natal disks}
\label{sec: prescription}

The main difference between the present investigation and that in
papers I-V are: 1) the inclusion of resonant capture 
between embryos during type I migration, and 2) the calculation of 
embryos' orbital and mass evolution after the gas depletion.  These 
effects are important for the formation of multiple Earths and 
super-Earth systems through collisional merger (giant impact) events. 
We describe in this section some modification and additions to our 
method of treating these interactions.

\subsection{Disk models}
\label{sec:diskmodel}
Although disk gas accretion rates in the inner regions and the surface
density of dust in the outermost regions of protostellar disks have
been observationally inferred, there are no reliable observational
constraints on the gas and heavy-element surface density distribution
especially near the planet forming regions. Theoretical determination
of these quantities relies on poorly determined magnitude of
angular momentum transfer efficiency. 

In light of this uncertainty, we adopt the MMSN model 
\citep{Hayashi81} as a fiducial set of initial conditions.  In
our population-synthesis model, we introduce multiplicative 
factors ($f_d$ and $f_g$) to scale the disk surface densities 
of gas ($\Sigma_g$) and planetesimals ($\Sigma_d$) with those 
of the MMSN model.   
Following Paper IV, we set 
\begin{equation}
\left\{ \begin{array}{ll} 
\Sigma_d & = \Sigma_{d,10} \eta_{\rm ice} 
f_d (r/ {\rm 10 AU})^{-q_d}, 
\label{eq:sigma_dust} \\ 
\Sigma_g & = \Sigma_{g,10} f_{g} (r/ {\rm 10 AU})^{-q_g},
\label{eq:sigma_gas}
\end{array} \right.
\end{equation}
where normalization factors $\Sigma_{d,10} = 0.32 {\rm g/cm}^2$ and
$\Sigma_{g,10} = 75 {\rm g/cm}^2$ correspond to 1.4 times of
$\Sigma_g$ and $\Sigma_d$ at 10AU of the MMSN model, and the step
function $\eta_{\rm ice} = 1$ inside the ice line at $a_{\rm ice}$
(eq.~[\ref{eq:a_ice}]) and 4.2 for $r > a_{\rm ice}$ [the latter can
be slightly smaller ($\sim 3.0$) \citep{Pollack94}].

In Paper V, we considered disk evolution with constant $\alpha$ and
the ice line barrier for migration (a local $\Sigma_g$ maximum is due
to a transition in the thickness of MRI active layers across the ice
line).  In this paper, we use a simple power-law disk model in order
to focus our attention on the effects of dynamical interactions that
we now take into consideration in our sequential planet formation model.
Nevertheless, we specify an inner disk boundary where $\Sigma_g$ vanishes
and planetesimals' type I migration is arrested (see \S2.5).

Neglecting the detailed energy balance in the disk \citep{Chiang97, 
Garaud07}, we adopt the equilibrium temperature distribution of 
optically thin disks prescribed by \citet{Hayashi81} such that
\begin{equation}
T = 280 \left(\frac{r}{1{\rm AU}}\right)^{-1/2}
    \left(\frac{L_*}{L_{\odot}}\right)^{1/4} {\rm K},
\label{eq:temp_dist}
\end{equation}
where $L_*$ and $L_{\odot}$ are stellar and solar luminosity.
In this simple prescription, we set the ice line to be that determined
by an equilibrium temperature (eq.~[\ref{eq:temp_dist}]) in optically
thin disk regions,
\begin{equation}
a_{\rm ice} = 2.7 (L_\ast/L_\odot)^{1/2} {\rm AU}.
\label{eq:a_ice}
\end{equation}
The magnitude of $a_{\rm ice}$ may be modified by the local viscous
dissipation \citep{Lecar06} and stellar irradiation \citep{Chiang97,
Garaud07}. The evolution of the ice line may affect frequency of gas
giant planets around high-mass stars \citep{Kennedy06,Kennedy08}.
These potentially important effects will be
incorporated in subsequent papers.

Dependence of disk metallicity is attributed to distribution of
$f_{d,0} = f_{g,0} 10^{{\rm [Fe/H]}_d}$, where $f_{d,0}$ and $f_{g,0}$
are initial values of $f_{d}$ and $f_{g}$, respectively. Due to
viscous diffusion and photoevaporation, $f_{g}$ decreases with time.
For simplicity, we adopt 
\begin{equation}
f_{g} = f_{g,0} \exp(-t/\tau_{\rm dep}),
\label{eq:gas_exp_decay}
\end{equation}
where $\tau_{\rm dep}$ is disk lifetime (for detailed discussion, see
Paper IV).  
The constant $\alpha$ self-similar solution obtained by
\citet{Lynden-Bell74} is expressed by
$\Sigma_g \propto r^{-1}$ with an asymptotic exponential
cut-off at radius $r_0$ of the maximum viscous couple.
In the region at $r < r_0$, $\Sigma_g$ decreases uniformly
independent of $r$ as the exponential decay does, although
the time dependence is different.
In the self-similar solution, $\Sigma_g$ at $r < r_0$ 
decays as $\Sigma_g \propto (t/\tau_{\rm dep} + 1)^{-3/2}$.
In the exponential decay model that we adopt,
$\Sigma_g$ decays more rapidly as
$t/\tau_{\rm dep}$ becomes larger at $t > \tau_{\rm dep}$.
If the effect of photoevaporation is taken into account,
$\Sigma_g$ decays rapidly after it is significantly depleted, so that
the exponential decay mimics the effect of photoevaporation.
We also carried out some runs 
with $\Sigma_g \propto (t/\tau_{\rm dep} + 1)^{-3/2}$,
but we did not find any significant difference 
in the results from the exponential decay cases.
In the population synthesis models in Papers I-V, a log
uniform distribution in a range of $10^6$--$10^7$ yrs was adopted for
$\tau_{\rm dep}$.  In the present paper, we fix the value 
$\tau_{\rm dep}$ to be $3 \times 10^6$ yrs.

\subsection{From Oligarchic Growth to Isolation}
\label{sec:oligar} 
On the basis of oligarchic growth model \citep{KI98,KI02}, growth rate
of embryos/cores at any location $a$ and time $t$ in the presence of
disk gas, is described by $dM_{\rm c}/dt = M_{\rm c}/\tau_{\rm c,acc}$
where, after correcting some typos in Paper IV, 
\begin{equation}
\tau_{\rm c,acc} = 2.2 \times 10^{5-q_d'-(2q_g'/5)} \eta_{\rm ice}^{-1}
f_d^{-1} f_{\rm g}^{-2/5} 
\left( \frac{a}{1{\rm AU}} \right)^{(27/10) + q_d'+(2q_g'/5)}
\left(\frac{M_{\rm c}}{M_{\oplus}} \right)^{1/3}
\left(\frac{M_\ast}{M_{\odot}} \right)^{-1/6}
{\rm yrs},
\label{eq:m_grow0}
\end{equation}
$M_{\rm c}$ is the mass of the embryo (core), $q_d' = q_d - 3/2$
and $q_g' = q_g - 3/2$, and we have adopted the mass of the typical
field planetesimals to be $m=10^{20}$g.  

Embryos' gravitational perturbations continually excite the eccentricity
of the field planetesimals.  Prior to the gas depletion, planetesimals'
eccentricity is also damped by hydrodynamic drag so that embryos' growth 
is enhanced by their gravitational focusing effect.  In this limit,
the full width of feeding zone of an embryo with a mass $M_{\rm c}$ is 
given by \citep{KI98,KI02}
\begin{equation}
\Delta a_{\rm c} = 10 r_{\rm H} = 
10 \left( \frac{2M_{\rm c}}{3M_{\ast}} \right)^{1/3}a, 
\label{eq:m_iso}
\end{equation}
where $r_{\rm H}$ is two-body Hill radius.  {In situ} growth of
embryos is quenched when they acquire all the heavy-element building
block material within their feeding zone. The maximum mass (called
isolation mass) of immobile embryos in oligarchic growth stage is
given by \citep{KI98,KI02}
\begin{equation}
M_{\rm c,iso} \simeq
0.16 \eta_{\rm ice}^{3/2} f_d^{3/2} 
\left(\frac{a}{1\mbox{AU}}\right)^{3/4-(3q_d'/2)} 
\left(\frac{M_\ast}{M_{\odot}} \right)^{-1/2} M_{\oplus}.
\label{eq:m_iso0}
\end{equation}

In Papers I-V, we adopted the above formulae for mass accretion
timescale during the oligarchic growth in the presence of disk gas. 
We also assumed that after the gas depletion embryos' and 
planetesimals' velocity dispersion becomes comparable to surface 
escape velocity of the most massive embryos at that location.
This kinematic transition widens embryos' 
feeding zone which enable 
them to acquire more planetesimals and attain larger asymptotic 
masses, though at slower rates (because their collisional cross 
section is reduced to their geometric surface area).  

In these previous first-attempt approximations, each embryo was
treated separately and their mutual interaction was neglected. For 
example, in Papers IV and V, decreases in $\Sigma_d$ due to accretion by 
independent planetary embryos are computed only in the feeding zones 
centered around their original locations. But, in the presence of disk 
gas, embryos undergo type I migration due to imbalance in the tidal
torques from outer and inner disks (see \S\ref{sec:migrat}). 
Relocation to regions with different surface densities (which is also
depleted by previous generation of embryos) would modify embryos' 
growth rate and asymptotic mass.  In order to take into account 
the possibility of growth along migration path, we adopted, in our 
previous papers, a prescription that embryos' growth rate is determined 
by the same value of $f_d$ in their original feeding zone until 
the migrating embryos reach the empirically-derived critical radius 
($a_{\rm dep,mig}$) within which the residual planetesimals are totally 
emptied by the preceding migrating embryos. This awkward scheme was 
introduced for computational convenience.  In the context of gas-giant 
formation, embryos evolve into dynamically isolated cores with nearly 
circular orbits in gaseous environment. This approximation is adequate 
for the evaluation of their individual growth and migration. It is also
reasonable to assume $\Sigma_d$ is depleted in this inside-out 
manner without the consideration of any competitive neighbors.

In the present paper, however, we intend to determine the mass
spectrum and kinematic distribution of multiple planetary
systems around common host stars.  Here, we need to consider 
the concurrent disk evolution along with the mass and dynamical 
evolution of several coexisting embryos in evolving disks (see 
below and \S\ref{sec:migrat}). Under some circumstances, it is possible
for the $\Sigma_d$ decline to be a non monotonic function of $a$. (For
example, the emergence of relatively massive embryos across the ice
line can lead to the formation of ``gaps'' in the $\Sigma_d$
distribution.)  In order to take this possibility into account, we
compute the evolution of $\Sigma_d$ distribution due to accretion by
all the emerging embryos in a self-consistent manner.

In our newly modified scheme, the growth and migration of several
embryos are integrated simultaneously with the evolution of the
$\Sigma_d$-distribution. We set up linear grids for $f_d$ across the
disk with typical width of $\sim 10^{-3}$AU. We introduce a population
of seed embryos, all with an initial mass $10^{20}$ g ({\it i.e.} that
of the residual planetesimals) and compute their mass accretion
rate. In the inner disk region (interior to the ice line), we set the
initial separation of these seed planetesimals at any given location
$a_0$ to be the full feeding-zone width ($ \Delta a_{\rm c} =10 r_{\rm
H}$) of embryos with local asymptotic isolation mass $M_{\rm c, iso}$.
We justify this prescription on the basis that in location where
$\tau_{\rm c, acc} (M_{\rm c, iso}) < \tau_{\rm dep}$, any seed
embryos with closer initial separations would have merged and
additional embryos would have formed unless the entire disk is filled
with isolated embryos.  Thus, this choice of seed planetesimals
is optimized for computational efficiency, without the loss of 
completeness. 

In the outer disk region where embryos' growth
is slow, they are unlikely to attain a local isolation masses
within the life span of their host stars.  A more realistic estimate
for their asymptotic mass is that ($M_9$) inferred from equation
(\ref{eq:m_grow0}) for the local $\Sigma_d$ after $t \sim 1$
Gyr. In the same spirit of optimizing computational efficiency
without introducing incompleteness,  we place seed embryos 
with an initial separation which corresponds to $\Delta a_{\rm c}$ 
(full feeding zone width) of embryos with mass $M_9$. For intermediate 
disk region, the initial 
spacing for the seed embryos is chosen to be $\Delta a_{\rm c}$ 
for the minimum value of $M_{\rm c, iso}$ and $M_9$. Provided there 
is a sufficient supply of seed planetesimals, our results do not
depend on the choice of their initial spacing.

We follow the growth of the seed embryos due to planetesimal accretion
in accordance with Equations (\ref{eq:m_grow0}) and (\ref{eq:m_iso}). 
The planetesimals' mass accreted by
the embryos is uniformly subtracted from the grids that are covered by
their instantaneous $\Delta a_{\rm c}$ appropriate for their current
mass and location. (For example, the feeding zone of an Earth-mass
planet at 1AU extends over 100 numerical grid points in the $\Sigma_d$
distribution.) We follow the evolution of $\Sigma_d$ in each
individual grids and use their values to evaluate embryos' accretion
rate and the strength of dynamical friction from the
planetesimals. With this treatment, we no longer need to assume some
empirical values for $\Sigma_d$ and $a_{\rm dep, mig}$ in the
evaluation of migrating embryos' accretion rate.

Embryos formed through oligarchic growth attain isolation mass and are
well-separated ($\sim \Delta a_{\rm c}$) from each other. Despite
their mutual gravitational perturbation, tidal drag from disk gas
\citep{Artymowicz93, Ward93} and dynamical friction from planetesimals
\citep[e.g.,][]{Stewart00} are sufficiently strong to generally preserve
embryos' circular orbits. Consequently, the stability of these widely
separated embryos is well preserved \citep{Iwasaki02}. In our
prescription, the process of embryos' oligarchic growth in a gas-rich
environment is adequately represented by planetesimal accretion onto
initially well-separated seed embryos. Nevertheless, embryos' mutual
perturbation may become important if their orbits evolve and masses grow
beyond isolation (see \S\ref{sec:postoli}).

\subsection{Type I migration}
\label{sec:migrat}
The growing embryos exert tidal torque on their natal disks.  With
sufficient masses, they undergo generally-inward type I migration
\citep{GT82,Ward86,Tanaka02}.  Several numerical
simulations have shown that type I migration can be retarded in
turbulent disks \citep{Laughlin04,Nelson05,Baruteau10} 
and weakly viscous or inviscid laminar disks
\citep{Koller03,Li09}.  The effect of
corotation torque and horseshoe drag can also reduce the rate and
change the direction of migration in disks with various $\Sigma_g$ and
$T$ distributions \citep{Masset09,Paardekooper10}. 
But, the
efficiency of these effects depends on the turbulent induced angular
momentum and heat diffusion coefficients.  

A detailed study of the type I migration is beyond the scope of this
paper.  However, we can utilize our population synthesis models to
calibrate the relative importance of several competing effects.  In
this paper, we follow the prescription in Paper IV and use the formula
for the migration rate that was derived through 3D linear calculation
\citep{Tanaka02}:
\begin{equation} 
\begin{array}{ll}
\tau_{\rm mig1} & 
{\displaystyle
= \frac{a}{\dot{a}} 
= \frac{1}{C_1}
  \frac{1}{2.728 + 1.082 q_g}
  \left(\frac{c_s}{a \Omega_{\rm K}}\right)^{2} 
  \frac{M_*}{M_p}
  \frac{M_*}{a^2 \Sigma_g}
  \Omega_{\rm K}^{-1} }\\
 & 
{\displaystyle
  \simeq 5 \times 10^4 \times 10^{-q_g'} \frac{1}{C_1 f_g} 
  \left(\frac{M_{\rm c}}{M_{\oplus}} \right)^{-1} 
  \left(\frac{a}{1{\rm AU}}\right)^{q_g} 
  \left(\frac{M_*}{M_{\odot}}\right)^{3/2}
  \;{\rm yrs}. } 
\end{array}
\label{eq:tau_mig1} 
\end{equation} 
where $C_1$ is a free parameter for the retardation of migration,
reflecting possible non-linear effects including turbulence (see Paper
IV). The expression of \citet{Tanaka02} corresponds to $C_1 = 1$, and
for slower migration, $C_1 < 1$. 
It also neglects the effect
of horseshoe torque associated with either $\Sigma_g$ and $T$
distribution.  The implication of these effects on the formation of
planets and planetary systems will be addressed in the next paper
together with that associated with a migration barrier at the inner
boundary of global dead zone.

In this paper, we are primarily interested in planetary systems
without gas giants. Equations (\ref{eq:m_grow0}) and (\ref{eq:m_iso0})
indicate that on a time scale $\sim \tau_{\rm dep}$, the most massive
embryos emerge from relatively low-mass disks (with $f_g = f_d \sim
1$) attain masses a few times that of the Earth.  Although these embryos
(or equivalently cores) can accrete gas, it forms slowly contracting
envelopes with insignificant masses (Paper I).  For
the case studies in \S3, we assume that prior to gas depletion, all
embryos have sub-critical masses $\la 3-10 M_\odot$ and neglect their 
gas accretion.

Orbital migration relocates embryos to regions with fresh supplies of
residual planetesimals. It can also lead embryos to regions where
planetesimals are severely depleted by the prior accretion of other
embryos. Equations (\ref{eq:m_grow0}) and (\ref{eq:m_iso}) are
formulae for oligarchic growth stage. They indicate that both $\dot
M_{\rm c}$ and $M_{\rm c, iso}$ depend on embryos' local $\Sigma_d$. In our
numerical scheme, we compute the instantaneous local $\dot M_c$ and
$M_{\rm c, iso}$ of migrating embryos together with the evolution of
$\Sigma_d$ (see \S\ref{sec:oligar}).

\subsection{Resonant capture}
\label{sec:resona}
In intermediate-mass disks (with $1 \la f_g \la 3$), some embryos form with
super-Earth masses.  Even with a ten-fold reduction in efficiency
(with $C_1 \sim 0.1$), these massive embryos undergo type-I migration
over significant distances from the ice line prior to the gas
depletion. Embryos' relocation also provides an opportunity to replenish 
their feeding zone and to grow beyond their initial isolation mass.  
Differential type I migration can reduce (as well as widen) the orbital 
separation between some embryos.  

Embryos with converging orbits may also capture each other into
mean-motion resonances.  After they enter resonances, these
embryos have a tendency to migrate together while maintaining the
ratio of their semimajor axes. N-body simulations \citep{McNeil,
Ogihara09} illustrate the possible formation of migrating convoys with
several resonant embryos.  In order to construct an utilitarian
prescription for resonant capture, we first compute the rate of type I
migration for each coexisting embryos separately and independently. We
identify convergent pairs ($i, j$) from their differential migration 
speed. Next, we consider dynamical perturbation between embryos of closest
pairs ($i, j$). Neglecting perturbation by other more distant embryos,
their orbital separation ($b = |a_i - a_j|$) expands impulsively after 
each conjunction. For nearly circular orbits, the expansion of embryos'
spacing after an encounter is given by a linear analysis
\citep{GT82,Hasegawa90} as
\begin{equation}
\delta b \simeq 30 
        \left( \frac{b}{r_{\rm H}}\right)^{-5} r_{\rm H},
\label{eq:del_b}
\end{equation}
where $r_{\rm H} = ((m_i + m_j)/3M_{\ast})^{1/3} a$. Since 
encounters occur at every synodic period [$T_{\rm syn} = 2\pi a/((d
\Omega/da)b) \simeq (4\pi a/3 b \Omega_{\rm K}$)], we find the
changing rate to be
\begin{equation}
\frac{d b}{dt} \simeq \frac{\delta b}{T_{\rm syn}} 
\simeq 7 \left( \frac{b}{r_{\rm H}}\right)^{-4} 
         \left( \frac{r_{\rm H}}{a} \right)^2 v_{\rm K}.
\label{eq:scat_mig}
\end{equation}

In the limit that $db/dt$ becomes comparable to the differential type 
I migration speed, 
$\Delta v_{\rm mig} = v_{{\rm mig},i} - v_{{\rm mig},j}$, 
convergent embryos' mutual interaction would be compensated by their 
relative motion (with speeds $v_{{\rm mig},i}$ 
and $v_{{\rm mig},j}$ 
respectively).  In this equilibrium, the separation of two convergent 
embryos would be maintained at  
\begin{equation}
b_{\rm trap} \simeq 0.16 \left( \frac{m_i + m_j}{M_{\oplus}}\right)^{1/6} 
         \left( \frac{\Delta v_{\rm mig}}{v_{\rm K}} \right)^{-1/4} r_{\rm H}.
\label{eq:b_trap}
\end{equation}
For computational simplicity, we set 
$v_{{\rm mig},i} = a/\tau_{{\rm mig},i}$ 
where $\tau_{{\rm mig},i}$ (for an embryos with a mass $m_i$) is given by 
equation (\ref{eq:tau_mig1}),
\begin{equation}
b_{\rm trap} \simeq 4 \times 10^{-q_g'/4} 
                    \left( \frac{m_i + m_j}{M_{\oplus}}\right)^{-1/12} 
                    f_g^{-1/4} C_1^{-1/4} 
                    \left( \frac{a}{1{\rm AU}} \right)^{q_g'/4} 
                    \left( \frac{M_{\ast}}{M_{\odot}} \right)^{1/4} 
                    r_{\rm H}.
\label{eq:b_trap2}
\end{equation}
In the above expression, dependences of $b_{\rm trap}$ on various parameters 
are very weak and its magnitude is always $\simeq 4 - 5 r_{\rm H}$. 

In the construction of a simple analytic approximation, we assume 
convergent embryos would capture each other into their low-order 
main-motion resonance with a semimajor axis separation comparable to
$b_{\rm trap}$.  We note that if the estimated value of $b_{\rm trap}$ 
is smaller than $2\sqrt{3} r_{\rm H}$, resonant capture would not 
actually occur, because embryos' separation is within their feeding 
zone and $\Delta b$ (in eq.~ \ref{eq:del_b}) is saturated due to 
the non-linear effect of overlapping resonances\citep[e.g.,][]{Nakazawa_Ida88}.

The condition for resonant capture is marginally satisfied for embryos
which undergo rapid type I migration with $C_1 \sim 1$. 
In this limit, trapping would be possible if multiple embryos have similar
masses and migration speed (so that their relative migration speed is
slow). N-body simulations \citep{McNeil,Ogihara09} show that several
embryos do migrate together in such cases.  However, near the disk inner
edge, earliest-generation embryos accumulate as their migration is 
halted there.  When subsequent-generation embryos approach these 
stalled population with their full migration speed, orbit crossing may 
occur because $b_{\rm trap}$ is reduced below $4-5 r_{\rm H}$.  After 
these embryos undergo dynamical relaxation and cohesive collisions
\citep{Terquem07,Ogihara09}, several embryos may eventually survive 
as merger products in some new resonance configuration. 

In our numerical scheme, we assume that in the limit of inefficient
type I migration (with $C_1 \simeq 0.1$), resonant trapping always
occurs at $b_{\rm trap} = 5 r_{\rm H}$. Even at the disks' inner edge, 
the separation between stalled resonant Earth-mass embryos is comparable 
to or larger than the total width of their feeding zones.  Resonant 
embryos which captured each other along their migration paths would 
spiral towards their host star in lock-step provided their differential 
tidal torque continues to enforce convergent orbital evolution. During 
subsequent migration, resonant embryos' eccentricities are excited 
due to the conservation of an adiabatic invariance \citep{Murray_Dermott}. 
But, they are also effectively damped by embryos' tidal interaction 
with the disk gas. Interior to the ice line in MMSN-type disks, an 
equilibrium is established in which the largest (Earth-mass) resonant embryos 
retain a small amount of non circular motion \citep{Chambers96,Zhou07}.  
In general, resonant embryos' equilibrium $e$ is less than the ratio 
of their separation to their semimajor axis $e_b \equiv b/a$ and 
the resonant embryos' orbits remain dynamically separated. 

Based on this consideration, we neglect the possibility of mergers 
between resonant embryos prior to gas depletion regardless whether 
they are migrating or stalled near the disks' inner edge. We assume
converging embryos capture each other into their low-order main-motion 
resonances for which their separation is $b_{\rm trap} \simeq 5 
r_{\rm H}$.  In our numerical scheme, we monitor the spacing between 
all migrating embryos. When $b$ between any pairs of resonant embryos is 
reduced below $b_{\rm trap}$ during orbital integration, we compute, 
for the next time step, their total angular momentum loss to the disk 
due to type I migration.  This loss is then redistributed among the 
resonant embryos such that they would migrate together with a 
fixed spacing $b = 5 r_{\rm H}$ between them.

When resonant embryos migrate to the inner boundary of the disk,
they endure a strong corotation torque from the disk gas which 
halts their orbital decay (see below). This migration barrier 
is maintained during the subsequent arrival and resonant 
capture of additional incoming embryos.


\subsection{Halting migration in the stellar proximity.}
\label{sec:halt}
In most inner disk regions (within $\sim$ 1 AU), super-Earth embryos
undergo rapid, inward, type I migration.  However, at special
locations where the gas surface density or entropy attain local
maxima, these embryos' type I migration may be stalled due to their
tidal interaction with the disk gas near their co-orbital region
\citep{Masset06,Paardekooper10}. In MMSN disks, a barrier against type
I migration is located at the inner boundary of the `dead zone' where
gas in the disk midplane is totally neutral and not directly affected
by magneto-rotational instabilities \citep{Kretke07}.  
However, this barrier moves inward and eventually vanish during 
the late stages of disk evolution when $\Sigma_g$ become sufficiently 
small for the entire disk to become active \citep{Kretke10}.  
Near both (outer and inner) boundaries of the dead zone, angular momentum 
transfer by MRI can induce local uniform rotation in the gas flow.
The outer part of these rigidly rotation zone attains
super-Keplerian flow which also induces a migration barrier \citep{Kato09}. 

Stellar magnetic torque can clear out disk gas inside the
radius where it is balanced by viscous stress in the disk 
\citep{Konigl91}.  In typical protostellar disks, 
this interaction induces the central stars to co-rotate 
with the Keplerian frequency at their inner edge.  
Type I migration may also be halted on the outer boundary of a
magnetospheric cavity \citep{Masset06}.
We discuss, in more detail below, a powerful halting 
mechanism at the outer boundary of the cavity. 

In contrast, gas giants induce gap formation in their natal disks 
and undergo type II migration which is generally inward inside a few
AU's \citep{LP85}.  Type II migration is generally 
stalled well inside the magnetospheric cavity. 
Since gas is severely depleted throughout this 
magnetospheric cavity, tidal torque between disks and planets 
well inside the cavity vanishes. 
In the limit of weak magnetic field, the size of the cavity ($r_m$) 
becomes comparable to or smaller than the stellar radii and 
gas giants may be halted by their tidal interaction with their 
rapidly rotating host stars \citep{Lin96}.  
Although gas giants are shielded and generally not affected 
by the stellar field directly,
it can potentially induce Roche-lobe overflow 
to halt their inward migration \citep{Laine08}.
Unless their host stars' magnetic and
spin axes are aligned, close-in gas giants would be exposed
to periodic field modulation and induced current in the upper
envelope on their night side where the magnetic diffusivity is
relatively large.  
Ohmic dissipation heats the envelope to induce their Roche-lobe overflow.
In either case, 
gas giants are likely to park closer to their host stars than super-Earths.

Thermal expansion of rocky embryos is generally negligible. 
However, rocky embryos generally have higher conductivity
than the atmosphere of their host stars so that the magnetic flux tube
connecting non-embedded planets and their host stars
would slip though the envelope of the host stars much faster
than across the planets. 
The potential drop across the field lines
drives a DC current which is analogous to that proposed for the
electrodynamics of the Io-Jupiter system 
\citep{Goldreich69}. The Ohmic dissipation of this current
produces a torque which drives the planet's orbit to evolve toward a
state of circularization and synchronization with the spin of the star. 
Around slowly (or rapidly) spinning stars, this process can also
cause rocky planets with periods less than a few days to undergo
orbital decay (or expansion/stagnation) within a few Myr,
which can affect the retention
efficiency of short-period super earths.
This effect is discussed in a separate paper \citep{Laine10}.
We neglect it in the present paper.

In Papers I-V, we artificially halted type II migration of gas giants
and type I migration of rocky/icy embryos at orbital periods of a few
days.  This prescription was introduced to simulate the effects of an
inner cavity in protostellar disks around magnetized T Tauri stars.
Using this prescription, we were able to infer a statistical estimate 
on the fraction of ``hot jupiters'' among all known gas giant planets.  
In this paper, we focus our investigation on the stalling mechanism
for super-Earths' type I migration at the outer radius ($r_m$) of 
the magnetospheric cavity.  Although the transition radius between 
the dead and active regions is also important for longer-period 
rocky/icy planets \citep{Kretke10}, we will consider this more 
complex effect in the next paper.

In order to specify a magnitude for $r_m$, we note that the observed 
spin periods of young stars show a bimodal distribution at peaks at 
about one week and one day \citep{Herbst05}
\footnote{
Recent Spitzer observations also support the bimodal
population \citep{reb06,CB07}, while Corot observation
may suggest single peak population, which means that
this issue is still controversial.
}.
\citet{Herbst05} suggested that for the 
one-week period stars the stellar magnetic field is coupled with the 
protostellar disk strongly enough to transfer spin angular momentum to 
the disk and open up a cavity, while the one-day period stars do not 
have a cavity. In order to take into account both possibilities,
we adopt two different inner boundary conditions: i) with or ii)
without cavities.

In models which neglect the presence of magnetospheric cavity, all the
embryos that migrate to a radius of 3-day orbital period (0.04AU in
the case of $M_*= 1M_{\odot}$) are removed from the sample under the
assumption they migrate all the way into their host stars without any
stoppage. In relatively massive disks, early-generation embryos grow
rapidly. Type I migration relocates these embryos from their birth
place and delivers them to their host stars before they can attain
isolation mass.  Nevertheless, subsequent generations of embryos 
may form out of the residual planetesimals.

In order to simulate this continuous formation and migration sequence,
we construct a prescription on the formation of next-generation seed
embryos. When type I migration of any embryo formed at $a_0$ has led
to the decay of its orbit to $0.5 a_0$, we inject a new seed embryo at
$a_0$. In contrast to the initial mass ($10^{20}$ g) for the 
first-generation seed embryos, 
the initial mass assigned to this born-again
embryo is chosen to be $10^{-2}$ times that of its predecessor.
If it is larger than the isolation mass estimated from
residual planetesimal surface density, the embryo mass
is given by the isolation mass. 
(This prescription is introduced to merely represent 
the mass growth of residual planetesimals.) 
The growth rate and asymptotic mass of this
embryo is determined by $\Sigma_d$ of the local residual
planetesimals.  For example, its growth would cease if the total mass
of planetesimals within its feeding zone is severely
depleted. Detailed prescriptions on when to introduce any seed embryo
and its assigned initial mass do not affect the final result.

In paper IV and V, we show that prior to severe gas depletion, this
self-elimination process reduces $\Sigma_d$ of the residual
planetesimals and the asymptotic mass of late-generation embryos.
During the gas depletion, a population of embryos is retained in the
disk. Many of these late-generation embryos have sufficiently low
masses to avoid extensive type I migration. In the absence of any
magnetospheric cavity, since all the early-generation embryos are lost
to their host stars, the mass and spatial distributions of the
asymptotically-retained embryos are not sensitive to the initial
magnitude of $\Sigma_d$. In the inner disk regions where 
growth time scale of embryos at their isolation masses 
is shorter than their migration time 
scale, they accrete all the planetesimals in their feeding zone.
After gas depletion, the retained embryos continue to perturb each 
other's orbits until they undergo orbit crossing and collide with 
each other (see \S\ref{sec:postoli}). In the outer disk region, embryos 
are embedded in residual planetesimals in their feeding zones. Through 
dynamical friction, these planetesimals damp embryos' eccentricities
even after the disk gas is completely cleared away 
\citep[][also see \S 3.2]{Goldrecih04}.  

Around strongly magnetized host stars, surface density of the gas
$\Sigma_g$ vanishes at $r < r_m$, where $r_m$ is the radius of the 
outer boundary of the magnetospheric cavity. Outside this cavity,
$\Sigma_g$ reaches a local maximum at $r = r_{\rm max}$.  In the 
zone at $r_m < r < r_{max}$, pressure gradient in the disk tends to 
be positive and angular momentum is transfered from the disk to the
isolated embryos through their unsaturated corotation resonances.  
Nevertheless, embryos may also lost angular momentum to the disk
through their Lindblad resonances \citep{Tanaka02,Tanaka04}. Embryos 
migration is quenched when they attain an torque equilibrium
\citep[][Paper V]{Masset06}.

When a pair of resonant embryos migrate to the inner disk edge, 
they experience a much stronger positive torque than that on a 
single embryo \citep{Ogihara10}. As we indicated above, resonant 
embryos' eccentricities are excited due to the conservation 
of an adiabatic invariance.  Asymmetric eccentricity damping
near the inner disk boundary leads to a net flow of angular 
momentum from the disk to the embryos. In the linear calculations 
\citep{Tanaka02,Tanaka04} the timescale of the eccentric damping 
($\tau_e$) is shorter than that of type I migration ($\tau_{\rm mig}$)
by a factor of $\sim (c_s/v_{\rm K})^2 \sim O(10^{-3})$. (In this paper,
we consider relatively slow migration with $C_1 \sim 0.1$ so that
$\tau_e/\tau_{\rm mig}$ is expected to be even smaller.)  Thus, angular 
momentum replenishment to the innermost resonant embryo is sufficient 
to compensate for the loss of it due to type I migration torques on 
all the trapped resonant embryos. Through a series of N-body simulations 
which include the damping due to disk-embryo interactions, 
\citet{Ogihara09} confirmed that this ``eccentricity trap'' can indeed 
stall the migration of a convoy of resonant planets. Based on these
results, we consider a limiting case in which the innermost embryos 
are halted beyond the edge of the cavity.

In the next section, we only present models in which type I migration 
is stalled at the outer boundary of the magnetospheric cavity. 
Nevertheless, we consider an alternative prescription in which we 
neglect the effect of eccentricity trap at the edge of the 
magnetospheric cavity.  In that case, innermost embryos are often 
forced into the cavity by the torque from outer embryos.  Models 
generated with this prescription predict a large population of very 
short-period planets which do not seems to be consistent with the 
findings of recent radial velocity surveys.  

For computational convenience, we set the edge of the magnetospheric
cavity at 0.04 AU. In reality, the location of inner edge depends on
the stellar magnetic field and gas accretion rate which evolve during
the disk depletion.  We will consider these evolutionary effects in
the next paper.

\subsection{Post-oligarchic growth after gas depletion.}
\label{sec:postoli}
In Papers I-V, we used Equation (\ref{eq:m_grow0}) to compute embryos'
planetesimal-accretion rate. This formula was accurate for the
population synthesis of gas giant planets because we need to determine
the formation and retention of sufficiently massive cores in gas-rich
environments (so that these cores can accrete gas and evolve into gas
giants).  In such an environment, cores' eccentricity is suppressed and
they attain isolation masses rather than engage in giant impacts.

Although their building-block embryos may have also formed in gas-rich
disks, rocky/icy planets' final assemblage need not proceed prior to
the gas removal. In typical protostellar disks, embryos' eccentricity
is effectively damped so that their growth is limited by dynamical
isolation. However, on the time scale of $\tau_{\rm dep} \sim$ a few
Myr, gas in these disks is depleted by viscous diffusion or
photoevaporation while residual planetesimals are exhausted by embryos'
accretion except in outer regions. 
With a decline in embryos' eccentricity damping efficiency,
their orbits become dynamically unstable on time scales which increase
with both embryos' separation and $\Sigma_g$
\citep{Chambers96,Zhou07}. Embryos' eccentricities
increase until their orbits cross (on a crossing time scale 
$\tau_{\rm cross}$) and their growth resumes through giant impacts. 
At the end of
this post-oligarchic growth, the masses of the largest embryos
typically increase by a factor of several to 30 \citep[e.g.,]
[]{Kokubo06}. Thus, giant impacts essentially determine the
asymptotic properties of rocky/icy planets.

Although we have not considered the dynamical interaction between multiple
embryos in Papers I-V, an effort was made to approximate the outcome of
giant impacts.  In our previous prescription, the possibility of mergers 
after the gas depletion was simulated with the expansion of embryos' 
feeding zone.  In the
construction of mass distribution for close-in rocky/icy planets
(Paper V), we also considered two extreme limits, {\it i.e.}  either all
or none of the embryos migrated to the stellar proximity merge.

In order to accomplish the task to simulate the mass, semimajor axis,
and eccentricity distributions of multiple super-Earths/Earths systems,
we need to construct a prescription which approximates the process of
embryos' eccentricity excitation and collisions through giant
impacts. Our previous application of Equation (\ref{eq:m_grow0}) to
the determination of mass growth associated with these giant impacts
are not accurate.  Here, we construct, in the Appendix, an improved
prescriptions for embryos' eccentricity growth in gas free environment
and giant impacts. We outline below our computational procedures in a
sequential order.

\noindent
1) We compile a list of both dynamically isolated embryos and
orbit-crossing pairs.  We evaluate the time scale ($\tau_{\rm
  cross}$, see Appendix) for all dynamically isolated embryos'
eccentricity to grow until they cross the orbits of their 
closest neighbors.

\noindent
2) We identify the pair of non-orbit-crossing embryos with the
shortest $\tau_{\rm cross}$.  

\noindent
3) After such a time interval has 
elapsed, we compute the expected statistical changes in their 
eccentricity and semimajor axis. We then identify all other
embryos whose orbits this pair would cross if these changes 
were implemented. 

\noindent
4) For this group of two or more embryos, we implement statistical 
changes in $e$ and $a$ due to repeated close scattering 
among themselves.

\noindent
5) We apply corrections on the magnitude of
semimajor axis changes among the participating embryos 
in order to preserve the conservation of total orbital energy.

\noindent
6) We identify pairs of impacting embryos based on their 
statistically weighted collisional probability. 

\noindent
7) Under the assumption that these events lead to cohesion,  
we adjust both $a$ and $e$ of the merger product to satisfy
the conservation of orbital angular momentum.

The search for potentially orbit-crossing pairs (step 2) are also 
applied to resonant embryos.  In this context, \citet{Terquem07} and
\citet{Ogihara09} found through N-body simulations that, for rapid
migration (with $C_1 =1$), several embryos remain locked in mutual
mean motion resonances near the disk inner edge after gas depletion
and dynamical relaxation.  These results indicate that multiple
resonant configuration with relatively small number of bodies is
stable and the formula for $\tau_{\rm cross}$ (eq.~[\ref
{eq:tau_cross} ]) cannot be accurately applied to such configuration.

However, \citet{Ogihara09} also found that for migration with 
$C_1 \ll 0.1$, resonant capture is more effective 
and produce convoys of
several (up to {\rm dozens}) resonant embryos over wide regions of the disk
ranging from its inner edge to radii beyond 0.1AU.  Resonant
configuration of these resonant convoys is dynamically unstable and
they always start orbit crossing and merge with each other after disk
gas depletion.  Since we are concerned with the relatively slow
migration ($C_1 < 1$) in this paper, we adopt eq.~(\ref{eq:tau_cross})
for $\tau_{\rm cross}$ even for the resonance-trapped bodies.

The above procedure is repeatedly applied while the number of residual
embryos declines and their separation increases with time. Eventually,
the magnitude of $\tau_{\rm cross}$ for all residual bodies exceeds
a Gyr age of their host stars. We classify these asymptotic
kinematic properties as dynamical architecture of mature systems.
Although these comprehensive procedures are complicated to integrate,
each step is based on well-studied celestial mechanics. Other than two
empirical parameters, there is no need to introduce any arbitrary
assumptions. The two parameters are also qualitatively inferred from
celestial mechanics (see Appendix), albeit their quantitative values
are calibrated by N-body simulations by \citet{Kokubo06}. Thus, this
semi-analytic scheme minimizes uncertainties in the embryos' dynamical
evolution.

\subsection{Comparisons between the semi-analytic and N-body simulations}
\label{sec:comparison}
In order to assess the validity of our semi-analytic scheme, we make
direct comparisons with the results of analogous N-body simulations.
\citet{Kokubo06} have performed N-body simulations for
the giant-impact stage from isolated embryos to terrestrial planets.
They systematically varied the initial disk mass, or equivalently the
isolation mass and number of embryos in the radial range of 0.5-1.5AU.
Since they carried out 20 runs with different initial azimuthal
distribution of the embryos for each disk mass, the results of these
simulations statistically represent asymptotic state of these systems.

\begin{figure}[btp]
  \epsscale{1.0}      
  \plotone{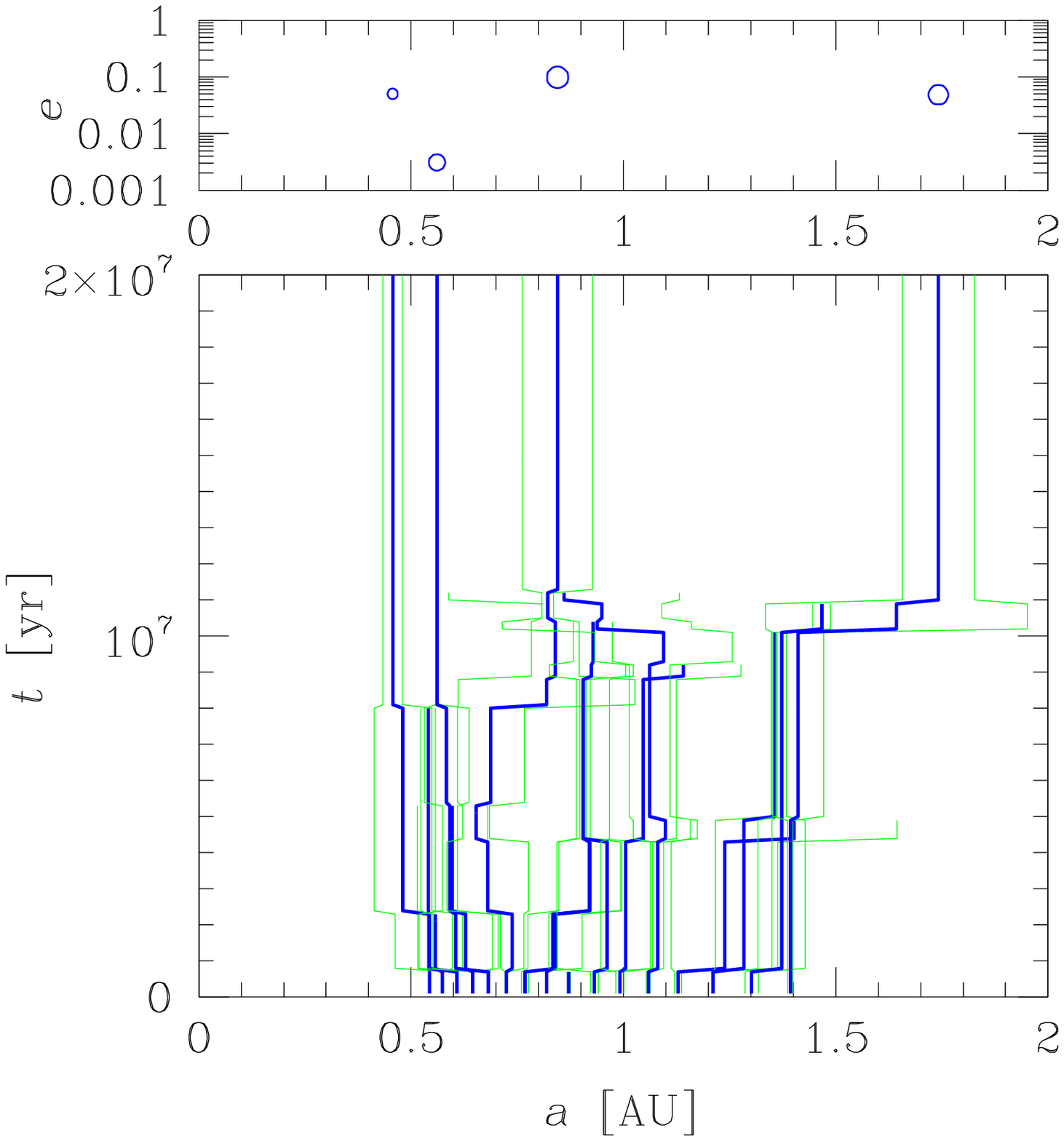}   
  \caption{   
An example of evolution of embryos in the post-oligarchic stage
calculated by the semi-analytical model. Initially, 16 embryos 
are placed in the disk with $f_d = 1$.  Thick and thin lines 
in the lower panel represent the evolution of embryos' semimajor 
axes and peri/apo-centers respectively. Discontinuities in lines 
represent merger events for embryos with others.  The upper panel 
shows semimajor axes and eccentricities of  final planets.
The radii of circles are proportional to physical sizes.  
}
  \label{fig:obt}
\end{figure}

With our prescription, we simulate the evolution of embryos systems
with similar sets of initial masses and radial distributions.  
Figures~\ref{fig:obt} show 
a typical example of embryos' post-oligarchic evolution.
In this model, we consider systems which initially contain 16 embryos 
with $M_p \sim $ 0.1--$0.2 M_{\oplus}$. These embryos represent oligarchics 
which have attained isolation masses in a MMSN-type disk ({\it i.e.,} 
with $f_{d,0} = 1$). The thick and thin lines in the lower panel of 
Figures~\ref{fig:obt} correspond to the semimajor axes and peri/apo-centers 
of the embryos' orbits.  Close scatterings lead to embryos' eccentricity
excitation and semimajor axis diffusion.  Line discontinuities 
represent merger events between nearby embryos.  The upper panel
shows semimajor axes and eccentricities of planets in an asymptotic
state.  All planets are assumed to have identical internal density 
and the radii of their representative circles are scaled to be 
proportional to their physical radii. In this model, four planets 
with masses $0.11 M_{\oplus}, 0.44 M_{\oplus}, 1.0 M_{\oplus},$ 
and $0.76 M_{\oplus}$ survive in stable orbits at 0.46AU, 0.56AU, 
0.85AU, and 1.74AU, respectively.

\begin{figure}[btp]
  \epsscale{1.0}        %
  \plotone{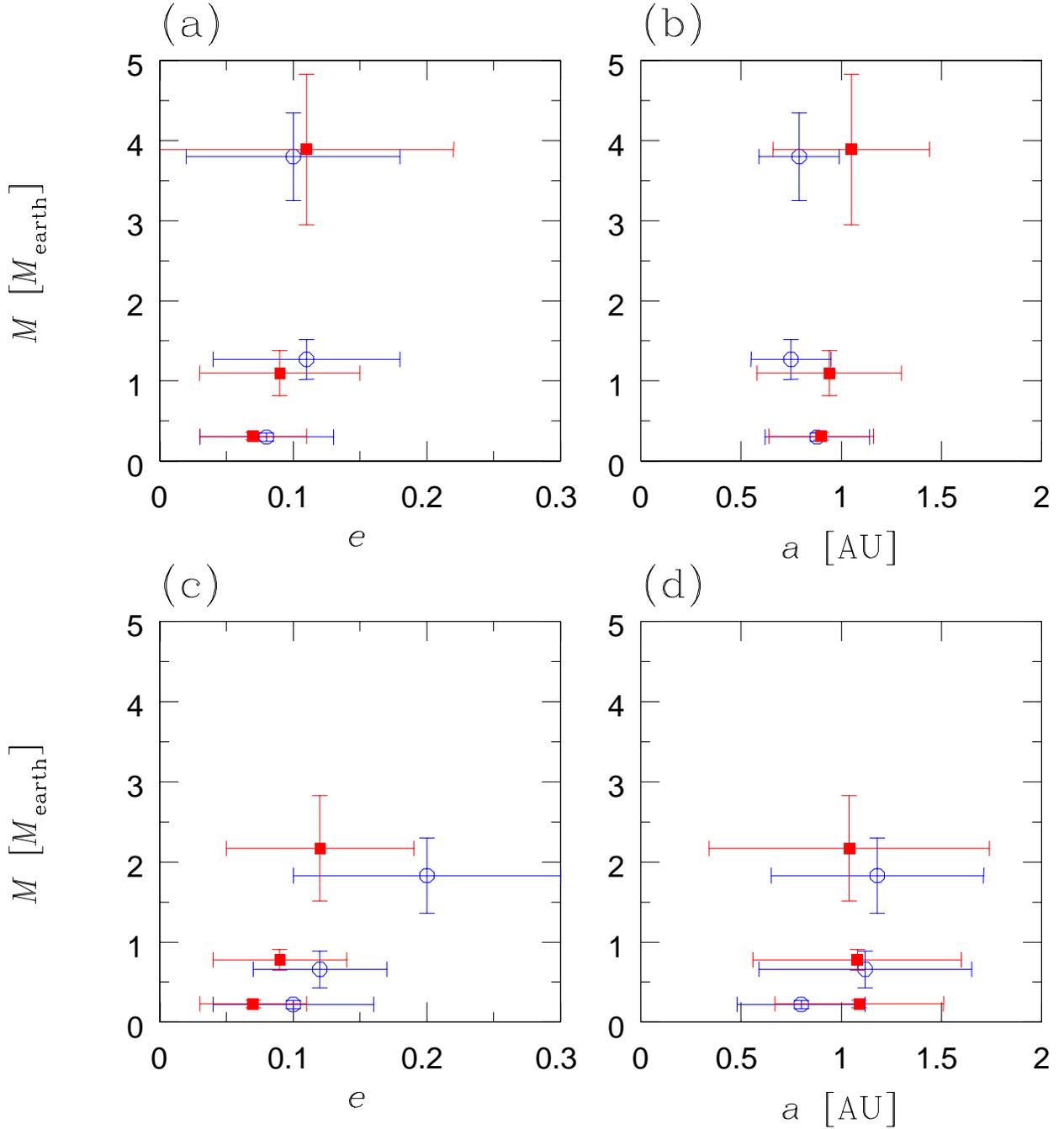}      %
  \caption{   
The averaged quantities of 20 models obtained by our semi-analytical 
model are represented with filled squares.  For comparison purpose, the 
results of N-body simulations of \citet{Kokubo06} are represented with 
open circles.  The bars indicate standard deviations.
(a) orbital eccentricity and mass of the most massive bodies,
(b) their semimajor axis and mass,
(c) orbital eccentricity and mass of the second most massive bodies,
and (d) their semimajor axis and mass.
the second most massive bodies (panels c and d).
Results for three sets of disk parameters ($f_d = 0.3, 1$ and 3) 
are plotted. The average masses increase with $f_d$.
}
  \label{fig:mae}
\end{figure}

In order to consider the statistical properties of these embryos
systems, we also simulated a set of 20 runs for the same
initial $f_{d}$ with our numerical
prescription, analogous to the previous direct N-body simulations.
Figures~\ref{fig:mae} show the averaged mass, semimajor axis, and
eccentricity of the most massive bodies and their standard deviations
in their final state.  For comparisons, these figures include the
results obtained by N-body simulations of \citet{Kokubo06} and our
semi-analytical model (panels a and b).  We also plotted the same
quantities for the second most massive bodies in panels c and d.  

The results plotted in these figures include sets of initial
conditions with $f_d = 0.3, 1$ and 3.  Embryos' initial masses are
given by their isolation masses and initial semimajor axes are
distributed from 0.55AU to 1.5AU with orbital separations of 
$10r_{\rm H}$.  The total mass of the distributed embryos is 
$0.72M_{\oplus}$, $2.3M_{\oplus}$, and $6.66M_{\oplus}$ 
for $f_d = 0.3, 1$, and 3,
respectively.  According to the values of $f_d$, number and masses of
initial embryos are changed.  We used the same initial conditions as
\citet{Kokubo06} (see their Table 1).

The figures show very good agreement in planets' asymptotic mass and
semimajor axis between the N-body simulations and our models for all
initial conditions.  The dependence of planets' asymptotic mass on
$f_d$ is well reproduced.  Although there is a difference in the
expectation value for $e$ of the second most massive bodies, it is
within a standard deviation.  Note that our model runs several orders
of magnitude faster than the direct N-body simulations, so we can
incorporate the post-oligarchic dynamical interactions into our
synthetic planet formation simulations.

\subsection{Transition from marginal metastability to protracted order} 
\label{sec:order}
After the embryos' eccentricity is excited to the magnitude which
enables them to cross other's orbit, their close encounters lead
to either elastic scattering or mergers. (For these self-gravitating
embryos, we neglect the effect of fragmentation and differentiation
between volatile and refractory materials.)  Repeated elastic
scattering further increases embryos' eccentricity to a value $e_{\rm
esc}$ which is the ratio of embryos' surface escape velocity and their
Keplerian velocity (see Appendix).  With this eccentricity, the
Safronov number is of the order unity and gravitational focusing no
longer dominates the collisional cross section.  

Orbit crossing events can also lead to direct collisions and merger
events.  These cohesive collisions reduce the number of surviving
planets and widen the separation between them.  Although some merged
embryos' orbits may become temporarily isolated, dynamical instability
can re-excite their eccentricity.  If this marginally stable state can
be maintained, the number of surviving massive embryos would continue
to decline while $\tau_{\rm cross}$ would become comparable to the
age of the system.

However, as we discuss in Appendix, $e$ of the merged embryos are
often significantly smaller than those of their progenitors prior to
the collision. These merger events generally involve embryos with
limited range of relative longitude of periastrons centered around 180
degrees. For Keplerian motion with $e \ll 1$, the mass-weighted total
Laplace-Runge-Lenz (LRL) vector is conserved \citep{Nakazawa_Ida88,
Nakazawa_Ida89}. Mergers of embryos with anti parallel longitudes of
periastron generally lead to fractional cancellation in the
mass-weighted total LRL and therefore relatively small values of $e$
of the merged embryos (for details, see Appendix). Consequently, the
asymptotic $e$ of the surviving embryos is generally smaller than
$e_{\rm esc}$. This efficient damping is consistent with the results
of previous N-body simulations \citep{Kokubo06}.

\begin{figure}[btp]
  \epsscale{1.0}        %
  \plotone{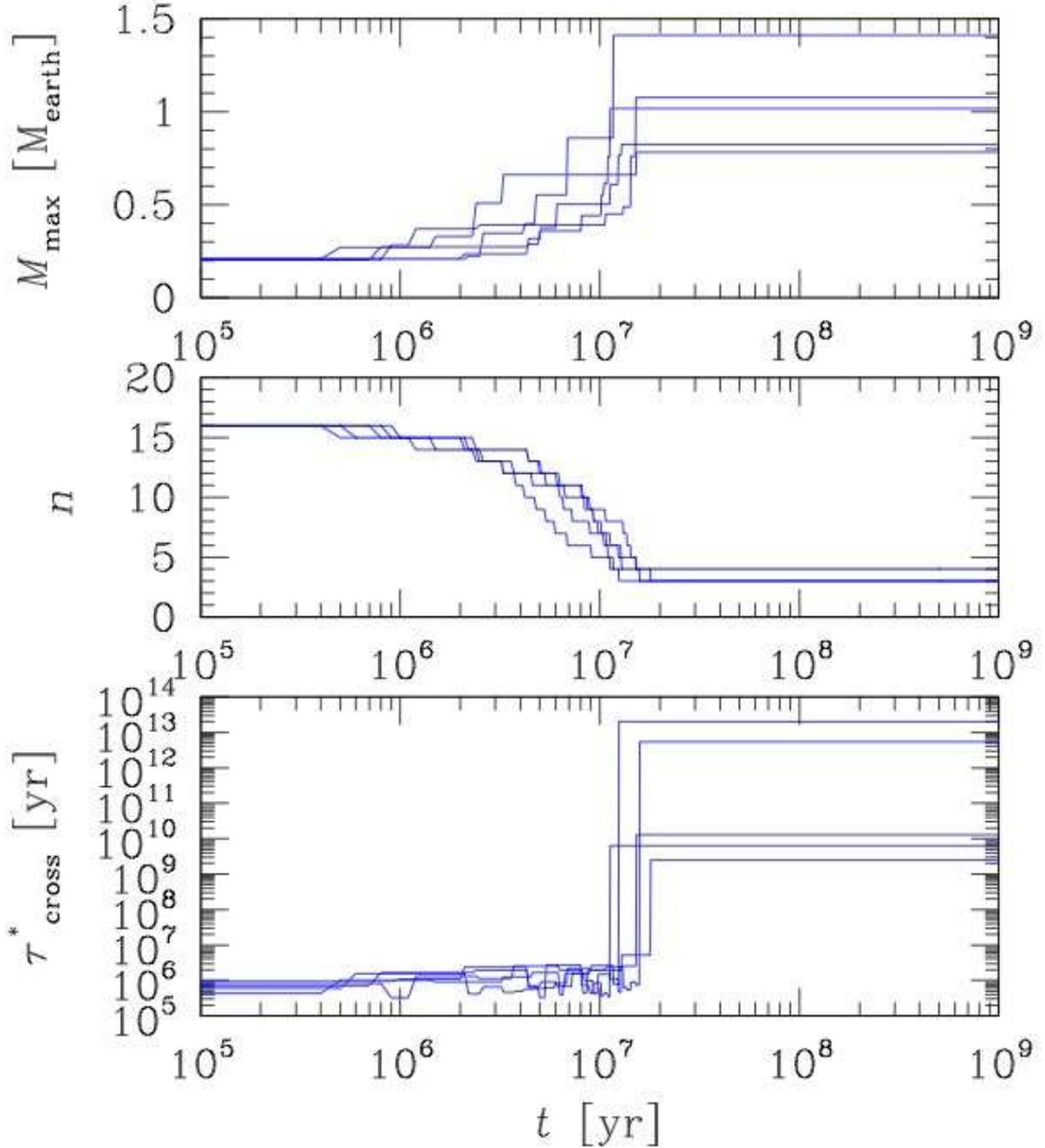}      %
  \caption{   
Time evolution of the mass of the most massive embryo
($M_{\rm max}$) is plotted on the top panel.  The total
number of residual embryos is plotted in the middle panel.
The minimum orbit crossing time for any pairs of embryos
is plotted in the bottom panel.  Five sets of models 
(generated with different random number seeds) for $f_d=1$
disks are presented here.  Each model starts with 16 embryos
at the onset of the simulation.
}
  \label{fig:t_cross}
\end{figure}

With this efficient damping process, many merged embryos' orbits
become temporarily isolated.  Although dynamical instability can
re-excite their eccentricity, the crossing time ($\tau_{\rm cross}$)
given by eq.~(\ref{eq:tau_cross}) often abruptly increases after some
merger events.  Figures~\ref{fig:t_cross} show time evolution of the
maximum planetary mass ($M_{\rm max}$) (the top panel), total number
(the middle panel) and the minimum crossing time (the bottom panel) in
five independent runs of our semi-analytical models with $f_d = 1$ and
initial number of the embryos $n = 16$.  After $n$ decreases to 3--4
and $M_{\rm max}$ grows to $\sim M_{\oplus}$, $\tau_{\rm cross}$
abruptly jumps from $\tau_{\rm cross} \la 10^{6}$ yrs to $\tau_{\rm
cross} \ga 10^{10}$ yrs, so the systems make a transition to a
dynamically stable state, on Gyr main-sequence life-span of solar-type
stars. Since these changes in the mass and eccentricity growth rates
are comparable to the gas depletion time scale in the disk, stochastic
merger events can introduce diversity in the extent of type I
migration of the surviving embryos.

\section{Population Synthesis of Planetary Systems}
With this new scheme, we simulate the formation of rocky and icy
planets.  The predicted mass, size, period, and eccentricity
distribution of close-in Earths/super-Earths that can be used to
directly compare with observational data \citep{Schlaufman10}.

\subsection{Initial conditions}
In Papers I-V, we presented a series of simulated planetary 
mass-semimajor axis distributions.  
We adopt a range of disk model parameters
which represent the observed distribution of disk properties and
assign them to each model with an appropriate statistical weight.
(For example, more massive ($f_g > 1$) and long lasting 
($\tau_{\rm dep} > 3 \times 10^6$ yrs) disks are adopted
less often than low-mass disks.)  For each disk, although several
planets may be generated over time, their mass growth and orbital
evolution are treated independently. (Evolution of the $\Sigma_d$
distribution was simulated with a simplified prescription, see
\S\ref{sec:oligar}). Mergers and giant impacts after gas depletion are
approximated by planetesimal-accretion formula for disks' outer regions
and artificially enforced (or neglected) in the stellar proximity.  
In Papers I-V, embryos' dynamical interaction with each other 
is neglected in
our population synthesis models, and planets' asymptotic
mass and period distributions are obtained from the compilation of
many monte carlo simulations. These properties only represent systems
which contain single planets, regardless of their masses.

Solar System contains four terrestrial planets, two gas giants and two
ice giants. Many extrasolar planets have known siblings.  Perhaps,
all planets reside in multiple-planet systems, albeit many members may
have sufficiently low mass to be below the detection limit.  
Nevertheless, their dynamical interaction may affect the overall
kinematic structure of planetary systems.  Even in systems which
contain only one massive gas giant, its dynamical influence on 
other residual embryos may affect their asymptotic architecture.

In this and subsequent papers, we will use our modified prescriptions 
described in the last section to generate statistically-weighted 
mass-period distributions for multiple-planet systems. Since there 
are many competing dynamical processes which may affect the outcome 
of these simulations, we adopt a step-by-step approach with 
increasing degrees of complexity.  Here we first carry out some 
case studies. In these preliminary models, we only consider 
systems around G dwarfs with solar mass and metallicity. (Stellar 
mass and metallicity dependences will be considered in future 
investigations.)  We focus in this paper on the formation of 
rocky/icy planets in the absence of any gas giants and
only consider disks with modest masses 
(with initial $ 0.3 \le f_d = f_g \le 3$).  
For these low-mass systems, planets' weak secular perturbation
on each other has been taken into account in our treatment of
post-oligarchic growth. (In addition to nonlinear close-encounters, gas
giants, if present, can also induce sweeping secular and resonant
perturbations which will also be considered in future papers.)

The initial distributions of $\Sigma_d$ and seed planetesimals are
specified in \S\ref{sec:oligar}. We introduce some small random
fluctuations to the initial locations of the seed planets for
different runs. Mass growth and orbital evolution of embryos are
computed with the prescriptions in 
\S\ref{sec:oligar}-\S\ref{sec:halt}. 
We do not consider the possibility of giant impacts among 
embryos in a gas-rich environment because these events must be 
preceded by embryos' orbit crossing (\S\ref{sec:postoli}). 
\citet{Iwasaki02} and \citet{Kominami02} showed that 
orbital crossing can 
occur only after $\Sigma_g$ is sufficiently depleted so that 
gas can no longer suppress embryos'
eccentricity excitation. Here, we adopt a necessary condition for
orbit crossing to be $f_g/f_{g,0} < 10^{-3}$.  Since we use $f_{g} =
f_{g,0} \exp(-t/\tau_{\rm dep})$ with $\tau_{\rm dep} = 3 \times 10^6$
yrs, orbit crossing is possible at $t > 2 \times 10^7$ yrs.  

Dynamical friction from a planetesimal swarm can also suppress orbit
crossing.  We calculate the total planetesimal masses in the feeding
zones of embryos at each timestep.  
If the total mass in their feed zones
is larger than any embryos' mass, we suppress their orbit crossing. 
(These embryos are not included in the orbit crossing bodies in step 
4 in the procedures
described in Appendix \S\ref{sec:postoli}).  Since their growth is slow 
in outer regions, embryos' perturbation on each other is limited and they 
do not undergo orbit crossing in the absence of any gas giants.
We assign an eccentricity $ \sim r_{\rm H}/a$ to each
embryo which does not undergo orbit crossing (see step 1 in Appendix).

\subsection{Overall evolution with dynamical interactions}
\begin{figure}[btp]
  \epsscale{1.0}       
  \plotone{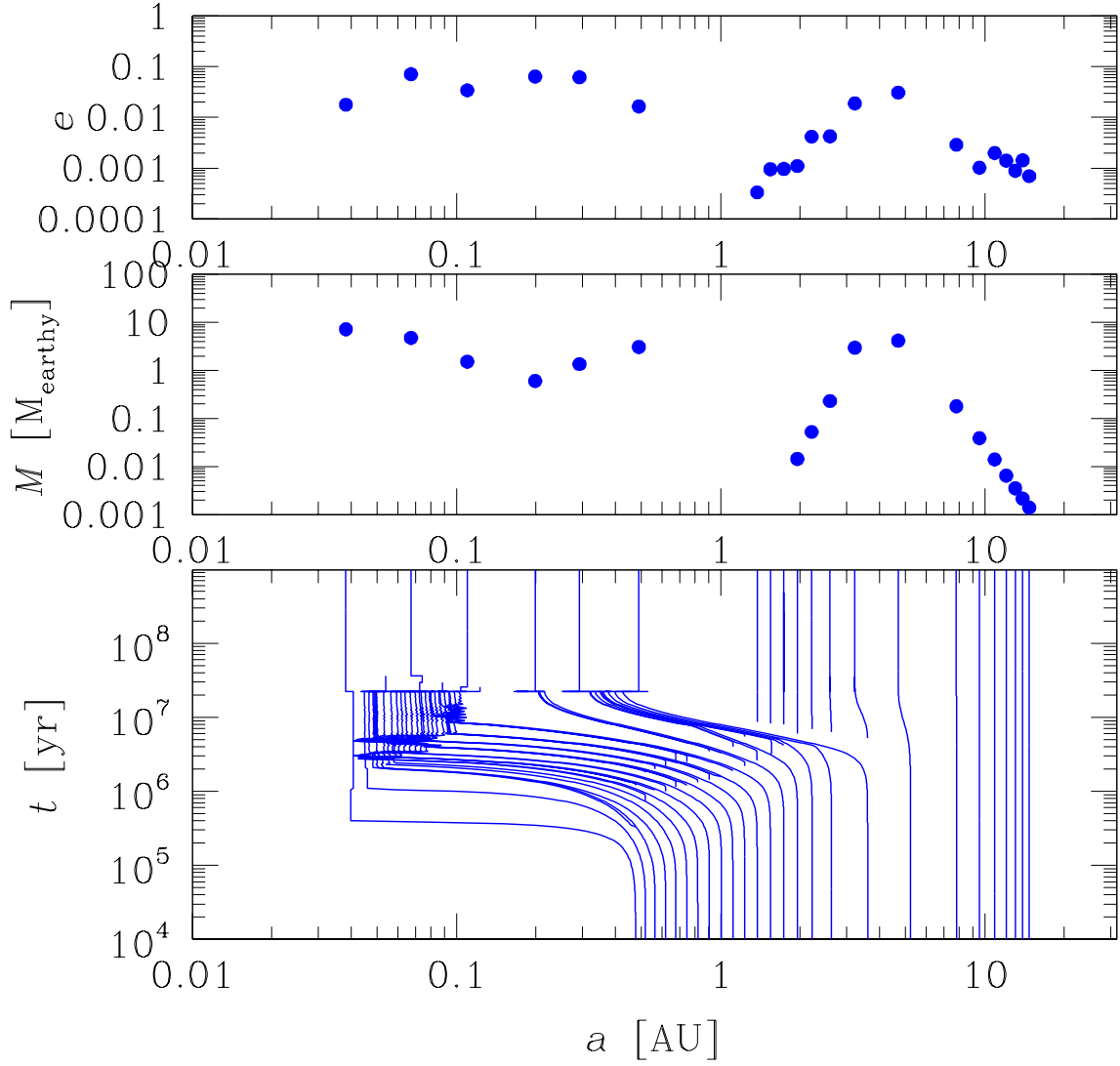} 
  \caption{   
Time evolution of semimajor axes of
all embryos is plotted in the bottom panel.  The asymptotic 
(at 1 Gyr) eccentricities and masses of all ``final'' planets 
are plotted in the top and middle panels. In this model, we
set $f_d = 2$. We also assume the presence of a disk cavity is
sufficient to stall the inward migration of all embryos in the
proximity of their host stars.
}
  \label{fig:obt_edge}
\end{figure}

We first consider a disk with a modest initial mass ($f_{d, 0} =2$)
and migration efficiency $C_1 = 0.1$.  For illustrative purpose, only
seed embryos at 0.5--15AU are integrated in this particular model.  
We apply two different sets of inner boundary conditions: 1) embryos' 
migration is stalled by a cavity (see Figures \ref{fig:obt_edge}) or 
2) they are not stalled at all.  (In subsequent papers, we will
consider the possibility that some embryos may be forced to undergo
further inward migration by the perturbation of additional embryos
which migrated to their outer low-order main-motion resonances.)
The bottom panels of these two figures show the time evolution of 
semimajor axes of all embryos and the top and middle panels show 
eccentricities and masses of final planets at $t = 10^9$ yrs.  

In inner regions, embryo growth due to planetesimal accretion and
migration are so fast that multiple-generation embryos are formed.  In
the bottom panels, the lines starting at $10^5$--$10^7$ yrs represent
2nd/3rd-generation embryos.  Due to their smaller masses, type I
migration of 2nd/3rd-generation embryos is relatively slow. They are
usually captured and shepherded by the mean-motion resonance of
1st-generation embryos that have migrated in from outer regions.  

At $t > 2 \times 10^7$ yrs, gas is sufficiently depleted that it can
no longer effectively damp embryos' eccentricity.  Thereafter, orbit
crossing and coagulations between embryos occur on timescales of
$10^7$--$10^8$ yrs.  In outer regions, embryos grow slowly and they do
not migrate over significant distances. Due to dynamical friction by a
large population of residual planetesimals, these embryos do not start
orbit crossing even at $t > 2 \times 10^7$ yrs.

In the model with a magnetospheric cavity (Fig.~\ref{fig:obt_edge}), a
convoy of migrating embryos (formed interior to $a \sim 4$AU) congregate 
and park near the disk edge. Their total mass is $\sim 20 M_\oplus$.
These embryos are captured into the mean-motion resonances of embryos
which have arrived in the stellar proximity at earlier times. This
group of resonant embryos extend from the inner edge of the disk to
radii beyond 0.1AU. (This phenomenon is observed in the N-body
simulation by \citet{Ogihara09} in their slow migration case.)  After
disk gas depletion, eccentricity excitation by embryos interaction
with each other is no longer damped by their tidal interaction with 
the disk.  As their eccentricity grows, embryos cross each other's
orbit.  Close encounters break up their mean motion resonance and 
this group of embryos collide with each other to form six 
planets with mass in the range of $\sim 1-10 M_\oplus$ between 0.03
and 0.5 AU.  In the absence of any residual gas, they cannot migrate 
into resonance again.

Figure~\ref{fig:sigma} shows the time evolution of the scaling factor
($f_d$) for planetesimal surface density due to accretion by embryos
in the case of the result in Figure~\ref{fig:obt_edge}.  In Paper IV,
we analytically derived the radius beyond which this surface density
is nearly preserved to be $a_{\rm dep,mig} \sim C_1^{1/8} (t/10^6{\rm
yrs})^{1/4}$ yrs.  That analytical formula agrees with the numerical
result presented here.  In outer regions, embryos start their migration 
when their masses are much smaller than their isolation masses, so the
depletion in $f_d$ there is small. These residual planetesimals
provide damping of embryos' eccentricities and thereby suppress orbit
crossing and giant impacts in their post-oligarchic stage.  However,
in inner regions ($\la 1$AU), most of planetesimals 
{\rm have been accreted by} 1 Myr.  
This clearing enables the post-oligarchic growth through orbit
crossing and coagulations between isolated embryos after gas
depletion.

\begin{figure}[btp]
  \epsscale{1.0}       
  \plotone{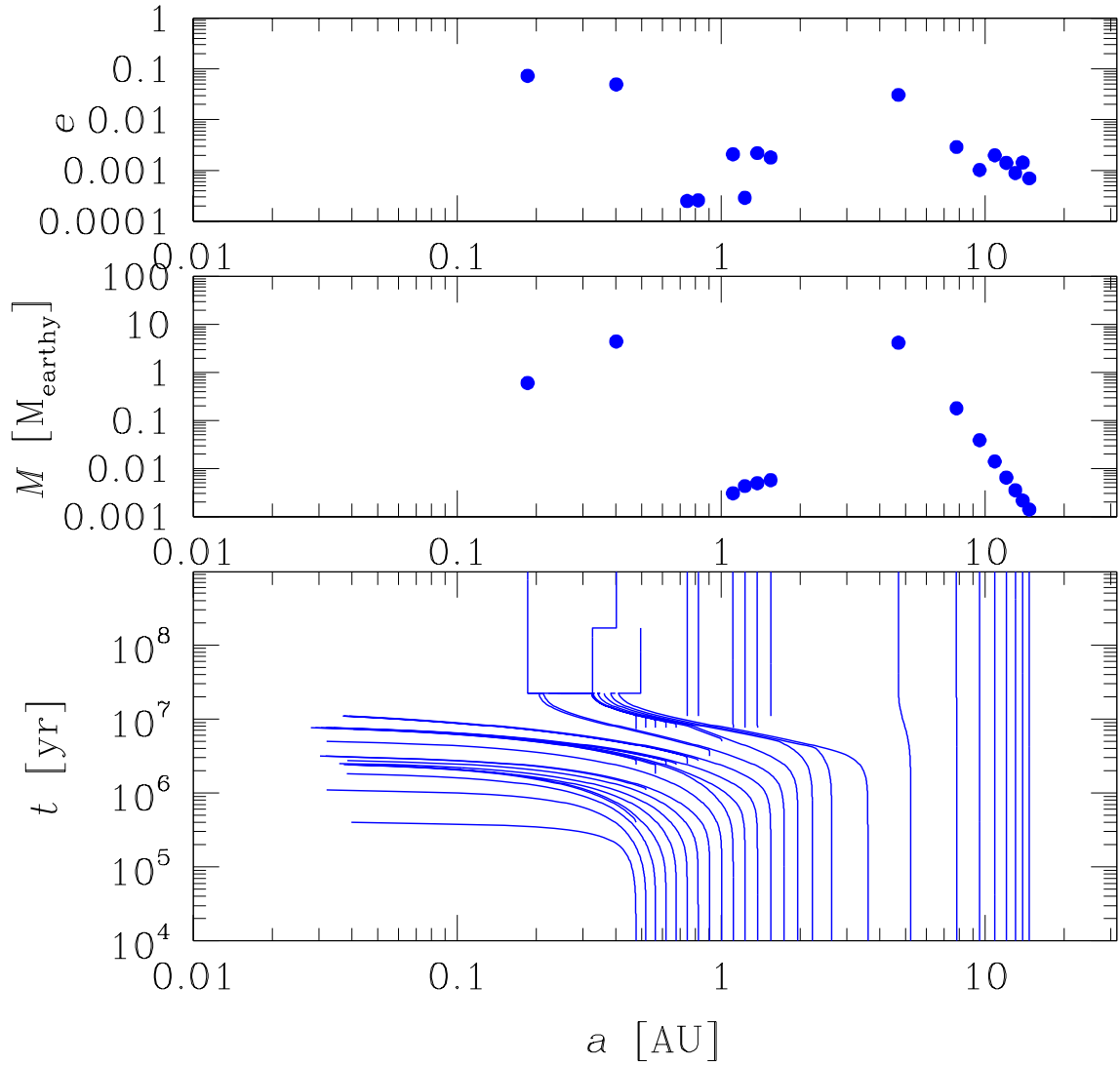} 
  \caption{   
The same as figure~\ref{fig:obt_noedge}
except we did not impose a migration barrier near the 
host star.  In our prescription, we adopt 
the no-cavity condition for the inner boundary.
}
  \label{fig:obt_noedge}
\end{figure}

\begin{figure}[btp]
  \epsscale{1.0}       
  \plotone{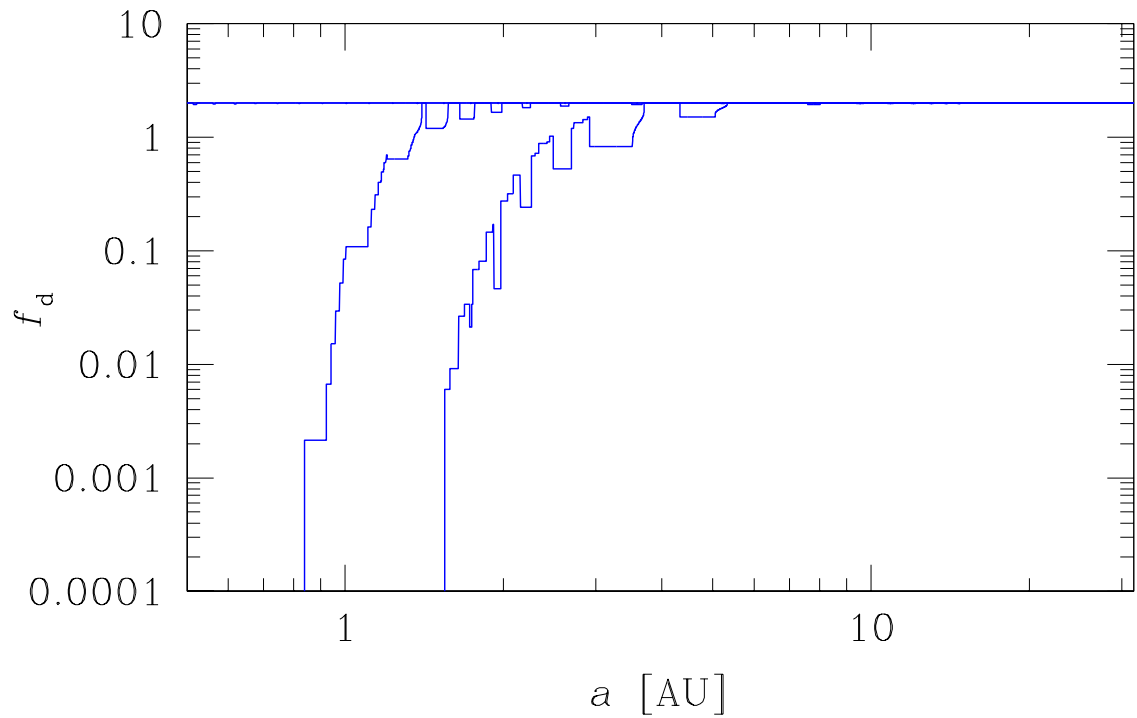} 
  \caption{   
The time evolution of planetesimal surface density
due to their accretion by embryos.  Same model parameters
are used as for the result in Figure~\ref{fig:obt_edge}.
The distributions at $t = 0, 10^4, 10^6$ and $10^8$ yrs are plotted
(the lines at $t = 0$ and $10^4$ yrs almost overlap with each other).
}
  \label{fig:sigma}
\end{figure}

We also consider a model without any migration barrier at the edge 
of the magnetospheric cavity (see Fig.~\ref{fig:obt_noedge}).  
In this case, many first generation embryos migrate 
into their host stars. 
No close-in ($\la 0.1$AU) Earths/super-Earths 
survives. Two super-Earths with $0.8 M_{\oplus}$ 
and $7 M_{\oplus}$ emerge at 0.18 AU and 0.4 AU respectively.

\subsection{Planetary growth in the inner and outer disk regions}
For a solar composition, 
the critical core mass to hydrodynamically sustain gas envelope is 
$M_{\rm c,hydro} \sim 10 M_\oplus$
\citep{Mizuno80,Pollack96,Ikoma00}.  Its magnitude also depends on
the planetesimal accretion rate, atmospheric composition 
\citep{Ikoma00} and
the boundary condition between protoplanets' atmosphere and their natal 
disk \citep{Ikoma01}.  In the stellar proximity where the 
disk gas is relatively dense and hot, magnitude of $M_{\rm c,hydro}$ 
may be smaller \citep{Ikoma01}. 
Once the core mass ($M_{\rm c}$) exceeds $M_{\rm c,hydro}$, 
envelope starts contracting.
If $M_{\rm c} > M_{\rm c,crit} \sim {\rm several}M_{\oplus}$, 
the contraction timescale
may be less than disk lifetime (Paper I), although
$M_{\rm c,crit}$ also depends on atmospheric composition.
If embryos arrive in the stellar proximity with 
$M_p \gg M_\oplus$, they may initiate rapid phase of gas 
accretion and evolve into hot Jupiters \citep{Bodenheimer00}.  

However, embryos' growth in inner regions is regulated by their 
type I migration and characterized by a two-stage growth 
(runaway/oligarchic growth and post-oligarchic growth) process.  
Prior to gas depletion, embryos with mass less than 
\begin{equation}
M_{\rm c,max} \simeq
0.21 C_1^{-3/4} 
\left(\frac{f_{g,0}}{3}\right)^{3/10}
\left(\frac{\eta_{\rm ice} f_{d,0}}{f_{g,0}}\right)^{3/4}
\left( \frac{a}{1{\rm AU}} \right)^{-9/8}
\left(\frac{M_\ast}{M_{\odot}} \right)^{5/4} M_{\oplus}
\label{eq:m_c_max}
\end{equation}
grow {\it in situ}, i.e., they gain mass faster than they undergo
type I migration.  (The critical mass for resistance against type I
migration $M_{\rm c,max}$ is determined by the condition $\tau_{\rm
mig1} = 3 \tau_{\rm c,acc}$, where a factor 3 reflects an actual
timescale to reach $M_{\rm c}$ because $\tau_{\rm c,acc} \propto
M_{\rm c}^{1/3}$.)  In regions of disks where embryos can attain
$M_{\rm c,max}$ before they acquire all the residual planetesimals 
within their feeding zone would migrate to and accumulate near
the disk inner edge. Prior to disk gas depletion, since these 
embryos cannot undergo orbit crossing, their growth through cohesive
collisions are temporarily quenched.  

Despite being surrounded by gas, these stranded embryos 
cannot evolve into gas giants in disks with modest 
$f_{d,0}(<10)$ because the magnitude of $M_{\rm c,max}$ is well 
below the critical core mass ($M_{\rm c,crit}$) 
for the onset of efficient gas accretion.  
Close-in embryos do significantly increase their masses 
through giant impacts to magnitude $>M_{\rm c, crit}$ 
after the disk gas depletion, especially in disks with 
$f_{d,0} \ga 2-3$.  But, there would be little residual
gas for these embryos to accrete and they are likely 
to evolve into super-Earths rather than the cores of gas 
giants.  Through this process, close-in super-Earths may 
bypass their isolation masses without becoming gas giants. 
Thus the detection of relatively massive compact 
planets (with $M_p > 10 M_\oplus$) does not necessarily 
imply a high magnitude for $M_{\rm c,crit}$.

In the outer regions, on the other hand, planetary growth 
ends in runaway/oligarchic growth stage and type I migration 
is much less effective. Since isolation mass increases 
with $a$, embryos emerge outside the ice line can acquire 
$M_p > M_{\rm c, core}$ and evolve into the cores of gas 
giants before gas depletion (Paper I). As we stated in the 
introduction, this paper focus on the low-mass disks which
do not produce gas giants.  In fact, in our disk models 
with $f_d = 2$, embryos with a few $M_\oplus$ emerge at 
$a\sim 3$--5AU (see Figs.~\ref{fig:obt_edge} and 
\ref{fig:obt_noedge}).  In the next paper, we will consider 
dynamical perturbation induced on the residual embryos by 
one or more emerging gas giants (we will also take into 
account the effect of an ice-line barrier in 
more massive disks, see Paper V).  
We anticipate a large fraction of residual 
long-period planetesimals and embryos may be ejected as 
they are scattered by one or more gas giants.

\subsection{Assemblage of rocky planets after gas depletion.}
\label{sec:rockyplanets}
In this section, we consider the detailed evolution of rocky/icy 
planets after gas depletion.  Using the prescription presented above, 
we present planets' asymptotic $M_p-a$ and $e-a$ distributions 
for several sets of simulations with a range of magnitude in $C_1$ 
and $f_{d, 0}$.  Seed embryos are initially distributed between
0.2--20AU (see \S 3.1).  Dynamical interaction between multiple
embryos is a stochastic process which would be inadequately 
represented by the results of any single set of simulations.  
In order to characterize planets' statistical distribution 
for each set of model parameters, 20 totally independent 
series of random numbers are used to compute their progenitors' 
dynamical interaction. From these Monte Carlo simulations, 
we obtain the mean values of $M_p$, $a$, and $e$ as well as 
their standard deviations. In this analysis, computation with 
our semi-analytical prescription is much faster (by several 
orders of magnitude) than direct N-body simulations.

\begin{figure}[btp]
  \epsscale{1.0}        %
  \plotone{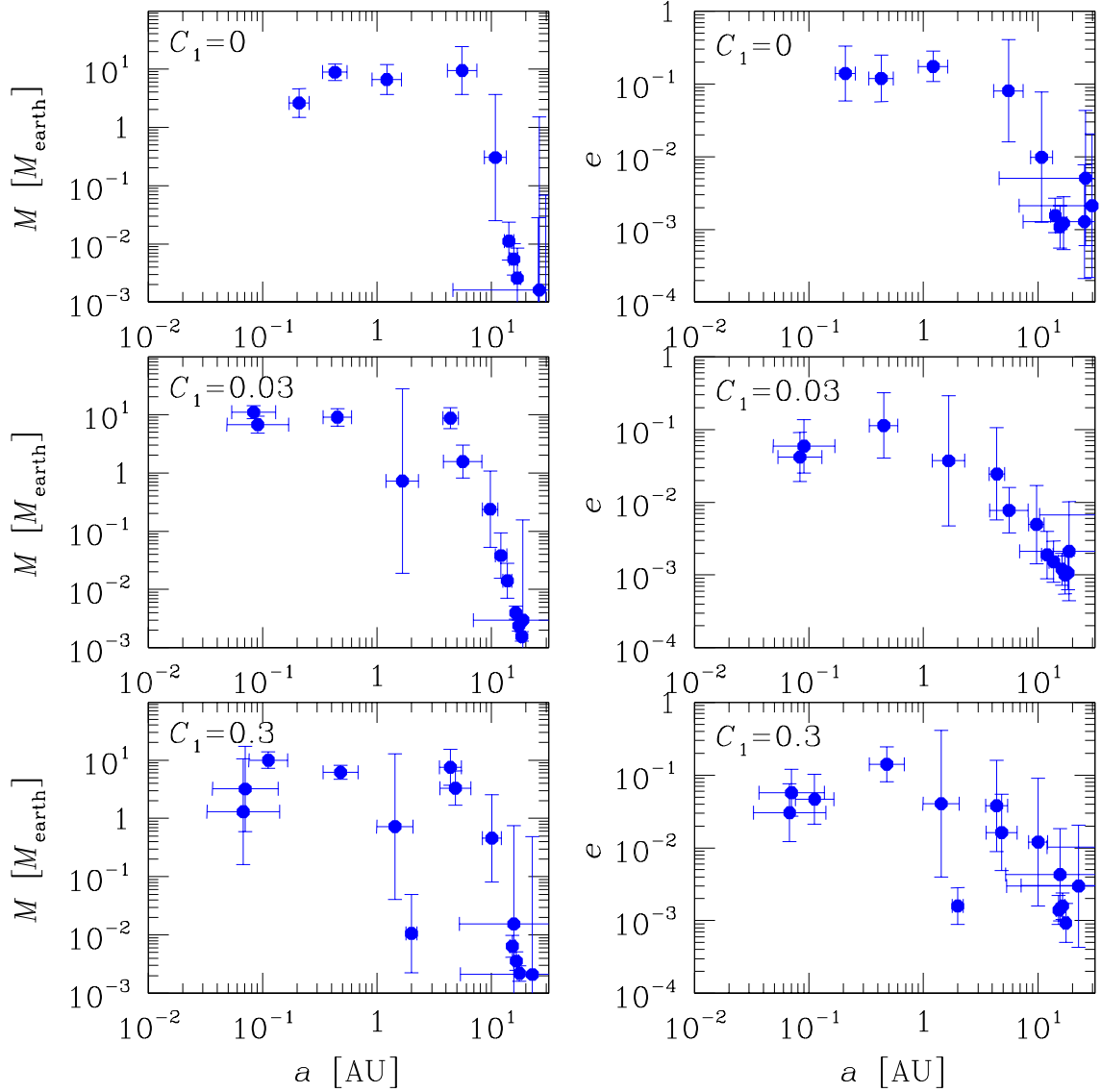}      %
  \caption{   
Simulated planetary systems for $C_1 = 0$, $0.03$, and $0.3$
are shown in the top, middle, and bottom panels respectively.
These results are generated from models with $f_{d,0} = 3$.
A migration barrier is imposed in accordance with the 
the cavity condition.  The left and right panels show asymptotic 
masses and eccentricities of bone fide planets at $t = 1$ Gyr.
The mean values of 20 runs are expressed by the symbols with
bars of standard deviations.
}
  \label{fig:stat_edge_f3}
\end{figure}

The asymptotic (at $t=1$Gyr) $M_p-a$ (left panels) and $e-a$ 
(right panels) distributions for models  with $f_{d, 0} =3$ 
and $C_1 = 0, 0.03,$ and $0.3$ are shown in the top, middle, 
and bottom panels of Figures~\ref{fig:stat_edge_f3}.  In 
these models, we consider highly magnetized host stars and 
impose a magnetic cavity in the disk and assume embryos' 
type I migration is stalled there (see \S\ref{sec:halt}). 
In \S\ref{sec:nocavity}, we consider the possibility of 
negligible stellar magnetic field.  

For presentation purpose, we divide the emerging embryos into
the close-in ($a < 0.1 a_{\rm ice}$), inner ($0.1 a_{\rm ice} 
< a < 0.3 a_{\rm ice}$), outer terrestrial ($0.3 a_{\rm ice} 
< a < a_{\rm ice}$), and icy ($a_{\rm ice} < a$) planet regions.
We record number of planets in each region, 
asymptotic (at $t = 1$ Gyr) mass, semimajor axis, 
and eccentricity of planets in order of mass in each region.
The quantities are averaged over the most massive planets, 
the second most massive planets,
the third most massive planets, ... in each region.
In Figures~\ref{fig:stat_edge_f3}, 
we plot planets of the averaged number in each region.

The results in Figures~\ref{fig:stat_edge_f3} show that 
typically three terrestrial (rocky) planets emerge in the 
$C_1 = 0$ (no type I migration) model. Since these planets
contain all the building blocks interior to the ice line,
they have masses of a few $M_\oplus$.  (In comparison, for 
the idealized model in Figure~\ref{fig:mae}, lower-mass 
planets emerge from planetesimals which were initially 
located within 1.5 AU in a MMSN.) These planets attain
relatively small orbital eccentricities ($\sim 0.1$). Their
corresponding velocity is considerably smaller than their
surface escape velocity. In \S 2.2 and Appendix, we suggest
that merger events generally lead to some degree of energy 
dissipation.  Nevertheless planets' asymptotic eccentricities
are larger than current eccentricities of Venus and Earth.
Dynamical friction from residual disk gas \citep{Kominami02}
or residual planetesimals \citep{O'Brien06} may further damp
the eccentricities.  Because we keep track the amount of 
residual disk gas and planetesimals, these effects can be 
accounted for in subsequent papers.

According to equation (\ref{eq:m_c_max}), Earth-mass rocky
embryos can relocate from all radii interior to the ice line
to the proximity of their host stars prior to the gas 
depletion even with inefficient type I migration. For 
models with $C_1 =0.03$, dozens of embryos form interior to
the ice line with masses $\sim M_{\rm c,max} \sim 0.2 
M_{\oplus}$ and then migrate to the vicinity of their host 
star.  After 
gas depletion, these embryos undergo dynamical relaxation
and cohesive collision to assemble into several super-Earths 
with $a \la 0.3$ AU.  Similar results are produced in models
with $C_1 =0.3$ (see middle and lower panels of 
Figures~\ref{fig:stat_edge_f3}).  
Although more efficient type I
migration depletes the residual planet-building blocks
at several AU, the asymptotic planet distribution close 
to their host stars is not very different between models 
with $C_1 = 0.03$ and $C_1 = 0.3$.  In both models, all
the migrated embryos are halted near the inner edge.
Post depletion giant impacts lead to the assemblage of
similar mass super-Earths, albeit the number of close-in
super-Earths appears to increase with $C_1$.

\begin{figure}[btp]
  \epsscale{1.0}        
  \plotone{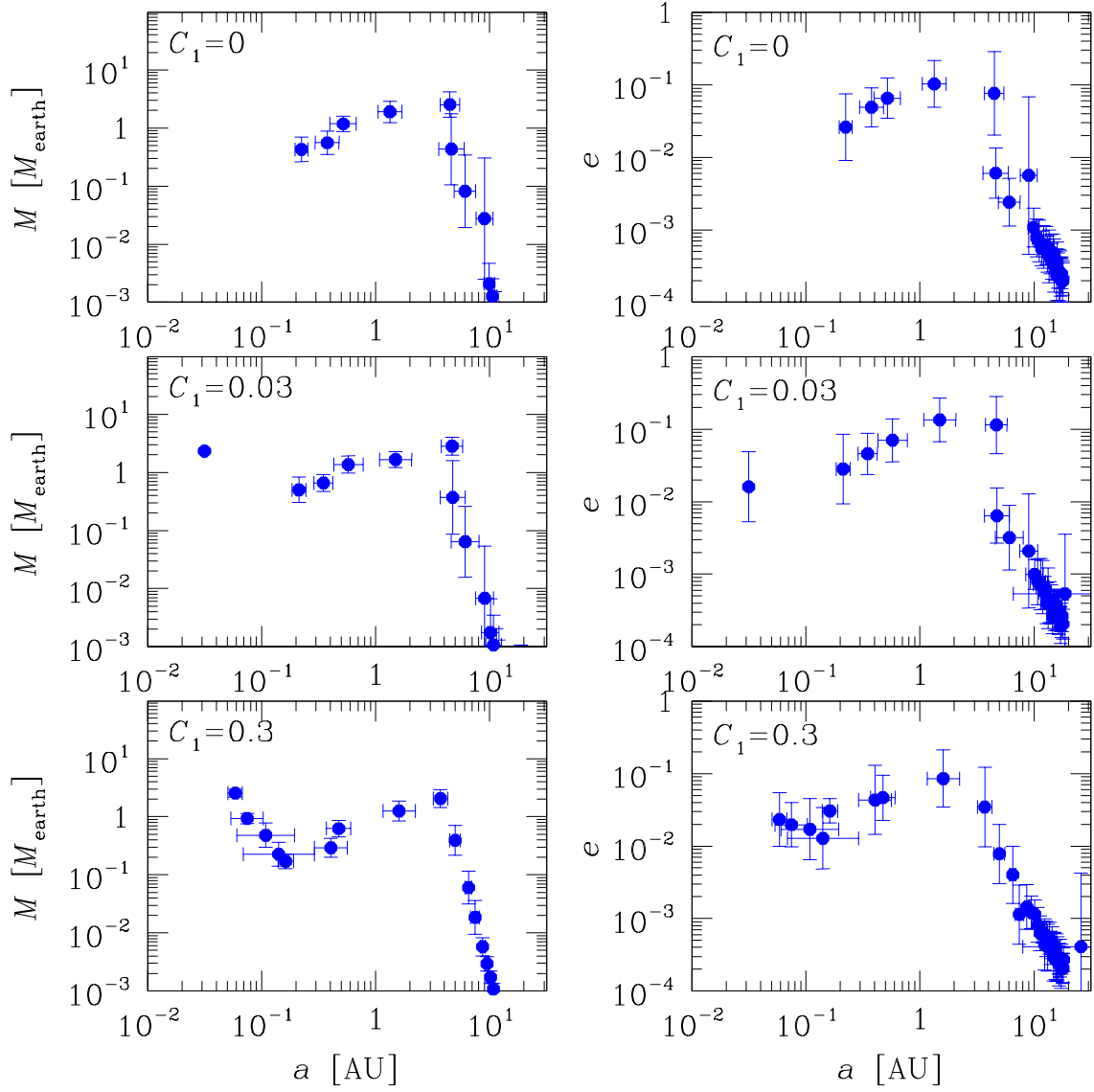} 
  \caption{   
The same as Figures~\ref{fig:stat_edge_f3} 
except for $f_d = 1$.
}
  \label{fig:stat_edge_f1}
\end{figure}

\begin{figure}[btp]
  \epsscale{1.0}        
  \plotone{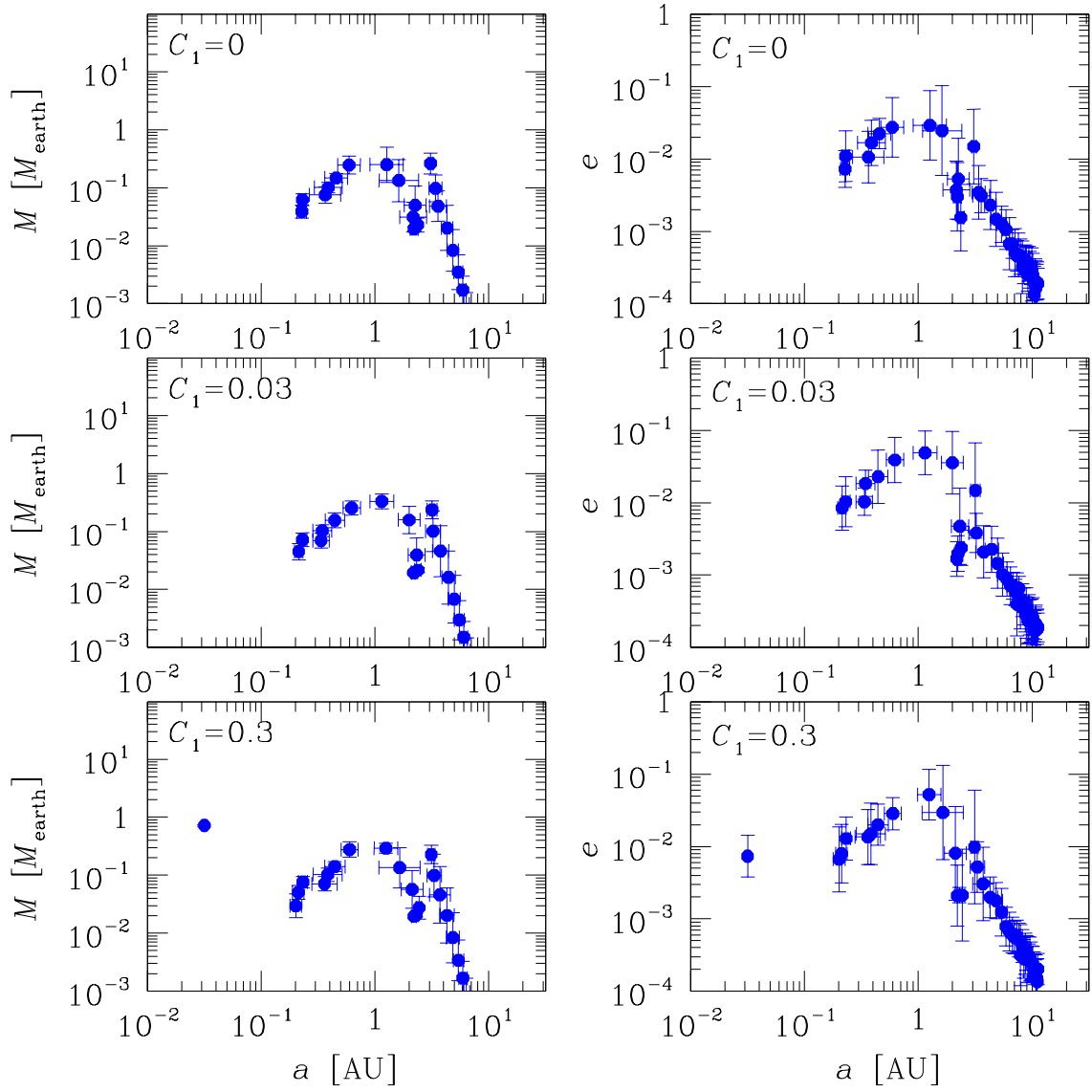} 
  \caption{   
The same as Figures~\ref{fig:stat_edge_f3} 
except for $f_d = 0.3$.
}
  \label{fig:stat_edge_f03}
\end{figure}

Simulation results for models with less massive disks 
are shown in Figures~\ref{fig:stat_edge_f1} 
($f_{d,0} = f_{g, 0} = 1$) and \ref{fig:stat_edge_f03} 
($f_{d,0} = f_{g, 0} = 0.3$).
Equation~(\ref{eq:m_c_max}) shows that in disks with
similar metallicity ($f_{d, 0}/f_{g, 0}$), 
$M_{\rm c,max}$
increases with $f_{g, 0}$.  In principle, small embryos can 
participate in type I migration.  But, embryos' growth is 
also limited by dynamical isolation.  Equation~(\ref{eq:m_iso})
indicates that the magnitude of the isolation mass
$M_{\rm c, iso}$ also increases with $f_{d, 0}^{3/2}$.  
In disk regions where $M_{\rm c, iso} < M_{\rm c, max}$,
many embryos emerge without type I migration.  In addition, 
Equation~(\ref{eq:tau_mig1}) 
indicates that the embryos' migration
timescale decreases with both their masses and $f_{g}$.
In the limit of inefficient ($C_1 = 0.03$) or no ($C_1 =0$)
migration,
embryos would not migrate over significant distance if their
$\tau_{\rm mig1} > \tau_{\rm dep}$.  Limited migration reduces 
the delivery of building block material to the stellar 
proximity. Consequently, the probability
of forming short-period super-Earths is an increasing function
of $f_{d, 0}$.  (The emergence of short-period Earth-mass planets 
from modest to low-mass disks requires relatively efficient type I
migration with $C_1 > 0.3$). 

In contrast, the retention of embryos 
near their cradles (just interior to the ice line) promotes the 
formation of rocky planets with intermediate
periods (months to years). After gas depletion, post-oligarchic 
growth continues through giant impacts on a time scale which is
an increasing function of $a$. In a MMSN-like disk (with $f_{d, 0}
= 1$), a system 4-5 planets emerge (on time scale of $\sim 
100$ Myr) with comparable masses, semimajor axes, and 
slightly larger eccentricities to those of the terrestrial 
planets in the solar system. Due to the difference in the
impact of type I migration, the formation probability of 
potentially habitable planets in this model is actually higher
than that out of more massive ($f_{d, 0} =3$) disks
(see Figures~\ref{fig:stat_edge_f3}).

\subsection{Formation of super-Earths around 
protostars with weak magnetic fields}
\label{sec:nocavity}

\begin{figure}[btp]
  \epsscale{1.0}        
  \plotone{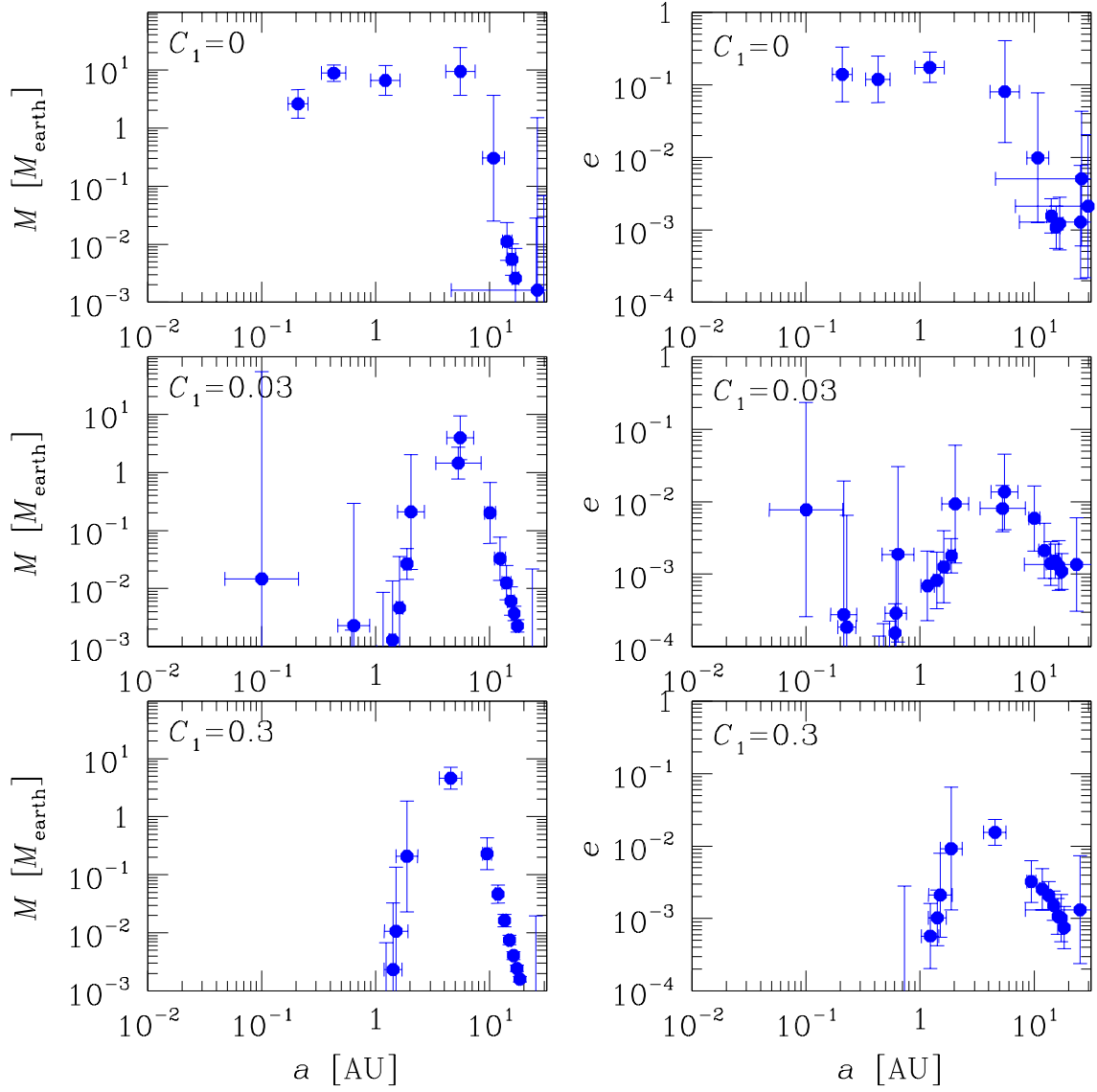} 
  \caption{   
The same as Figures~\ref{fig:stat_edge_f3} 
except for the non-cavity condition.
}
  \label{fig:stat_noedge_f3}
\end{figure}

In the limit of negligible stellar magnetic field, protostellar
disks extend to the stellar surface (see \S\ref{sec:halt}).  
In the absence of a magnetospheric cavity, migrating embryos 
are unable to be halted.  The importance of this inner boundary
condition is highlighted in Figures~\ref{fig:stat_noedge_f3} 
which represent models with the same value of $f_{d, 0} (=3)$ 
as the model in Figures~\ref{fig:stat_edge_f3}. (In this series
of simulations, the onset of efficient gas accretion is artificially
suppressed.  In a subsequent paper, we will consider the 
concurrent formation of gas giants and super-Earths.)

With modest migration efficiencies ($C_1 = 0.03$ and 0.3), embryos 
formed within $\sim 1-2$ AU prior to gas depletion attain mass 
$\sim M_{\rm c,max}$ and migrate all the way into their host
stars. They either {\rm accrete residual planetesimals
or capture other embryos} along their migration paths and 
clear the inner ($\la 1$ AU) region around their host stars.  
At $a \ga 2$AU, embryos' migration is initiated before they 
reach their isolation masses (see Paper IV). After the gas depletion,
multi-generation embryos with non-negligible masses are retained. 
These embryos undergo orbit crossing as their eccentricity is 
excited by their mutual perturbations. Through post-gas-depletion
giant impacts, they merge into Earth-mass planets. Beyond the ice line, 
cores at 4--5 AU have masses of $\sim 10 M_{\oplus}$. But, these systems
do not contain any Earth-mass planets with periods less than a few month. 
This asymptotic dynamical architecture ({\it i.e.}, $M_p-a$ distribution) 
is similar to that of the Solar System.

\subsection{Production of debris disks}
Observationally, there is no clear correlation between 
the detection of gas giant planets and that of debris 
disks around common host stars.  While the fraction of
solar type stars with known gas giants increases with
their metallicity \citep[e.g.,][also see Paper II]{Fischer05}, 
there is no analogous 
correlation between stellar metallicity and the 
presence of debris disks around them 
\citep{Greaves}.

Debris disks are composed of grains with sizes comparable
to the infrared wavelengths (a few $\mu$m to sub-millimeter).
Around mature solar stars, smallest (sub-micron) grains are 
blown out by radiation pressure. Due to Pointing-Robertson 
drag, modest (sub-mm) grains undergo orbital decay on time 
scale (a few Myr) shorter than the age of their host stars 
\citep[e.g.,][]{Takeuchi01}.  Thus, these grains must be
continually generated through collisions of their parent 
bodies.  

In all the models we have considered here (regardless the 
magnitude of $C_1$), embryos growth prior to gas depletion 
is limited by dynamical isolation in the inner disk regions.
After gas depletion, post-oligarchic giant impacts inevitably
occur, regardless of the stellar metallicity.  Thus, the 
production of ``warm'' debris dust (with wavelength up to
$\sim 10 \mu m$) is expected to be common around all
solar type stars.  

If, in the inner disk region,  a large fraction of the total 
mass in heavy elements is attained by the largest embryos, 
time scale generating a large pool of warm debris particles 
($\tau_{\rm cross}$) would be longer than the grains' clearing 
time scale but shorter 
than the age of mature host stars (Gyr). Thus, replenishment 
of detectable grains in debris disks may be a stochastic 
process and the intensity of debris disk signature may vary 
episodically \citep{Kenyon_Bromley04}.

In outer disk regions, planetesimal growth is incomplete
with a large amount of residual planetesimals (see \S 3.2).
Embryos' eccentricities in this region are likely to 
be excited by close encounters or merger events with other 
embryos and damped by gas through tidal interaction and 
residual planetesimals through dynamical friction. Unless
icy embryos become sufficiently massive to efficiently 
accretion gas (more massive than $M_{\rm c, crit}$), their
surface escape velocity ($\sim 10$ km s$^{-1}$) is generally 
smaller than Keplerian velocity at $a < 10$ AU.  Consequently,
close-encounters with embryos do not lead to the ejection of
residual planetesimals. In this limit, planetesimals' velocity 
dispersion is excited by embryos' repeated gravitational
perturbation and damped by inelastic collisions among 
themselves. When an equilibrium is established, planetesimals' 
velocity dispersion becomes comparable to the surface escape 
speed of the planetesimal population which carries most of the 
masses.  

If these planetesimals have sizes larger than 
a few km's (as in the Kuiper Belt, 
\citep[e.g.,]{Pan05}, 
their velocity dispersion would exceed that 
for disruptive collisions \citep{Stewart09}.
Debris of such collisions would produce ``cold'' grains which
reprocess stellar radiation and generate far infrared excess
radiation \citep{Wyatt08}. Since the collision time scale for
numerous km-size planetesimals is expected to be shorter than
the grain depletion time scale, the signature of outer debris
disk is expected to be maintained at a steady level.

In relatively massive disks where one or more embryos can 
attain critical mass for the onset of efficient gas accretion
prior to the depletion of the disk gas, the formation of gas
giants can scatter residual planetesimals to large distances.
For example, in the Solar system, while residual planetesimals 
near Uranus and Neptune were scattered by them to the Kuiper
Belt region or the Oort's cloud, 
those in the vicinity of Jupiter and Saturn were
either ejected or scattered to the distant Oort's clouds
\citep[][]{Duncan97}.
The clearing of residual planetesimals (as parent bodies) 
would suppress the signature of cold debris disks around stars 
with gas giants.  Thus, around host stars with relatively low
metallicity, the lack of gas giants does not necessarily
imply a lower detection probability for debris disks.

\subsection{Formation of close-in super-Earths}

The results in previous sections indicate that a 
small amount of planet-disk tidal interaction can lead 
to significant type I migration for embryos more massive
than $M_{\rm c, max}$.  In sufficiently massive disks
(with $f_{d,0} > 1$), a convoy of embryos stall from 
outer edge of magnetospheric cavity (specified to be
0.04 AU in our models) to locations beyond 0.1 AU.
(Around strongly magnetized stars or in less massive
disks, embryos may be stalled at larger distances from
their host stars). These embryos merge through giant 
impacts after gas depletion.  Since only a fraction of 
embryos' energy is dissipated, the resultant semi 
major axis of the merger products is expected to be 
comparable to that of their progenitor embryos.  

Close encounters between embryos generally
do not lead to mean motion resonance. In the absence 
of residual disk gas, these merger products do not 
undergo any further orbital decay and generally remain
out of mean motion resonance with each other. These 
expectations are consistent with the simulated asymptotic
$M_p-a$ distribution (see Figs.~\ref{fig:stat_edge_f3} 
and~\ref{fig:stat_edge_f1}). 

The velocity dispersion of the residual embryos is a 
fraction of their surface escape speed.  In the stellar
proximity, it is much smaller than the local Keplerian 
speed.  In comparison with Earth-mass planets at around 
1AU, the simulated eccentricities of close-in super-Earths 
are relatively small (in the range of 0.01--0.1). In
subsequent papers, we will consider the possibility of
excitation of much larger eccentricity by secular resonances
of distant gas giants 
\citep[e.g.,][]{Mardling_Lin2004, Nagasawaetal05, 
Thommesetal08}.  
 
Many emerging planets have masses larger than $\sim 
M_{\rm c,crit}$ which is the critical mass for cores
to evolve into gas giants within a few Myr.  Since
these planets acquired most of their mass after the 
gas is depleted or in gas-free cavities, they cannot
accrete a substantial gaseous envelope. The atmosphere
of their progenitor embryos may also be ejected during
giant impacts.  Despite their challenges in attaining
primary atmospheres, the emerged super-Earths may attain
a metal-rich secondary atmosphere through outgassing.  
However, such an atmosphere would contribute to a small
fraction of these planets' total mass.

\subsection{Frequency of habitable planets}
In disks with $\Sigma_g$ and $\Sigma_d$ comparable to those 
of MMSN, embryos' isolation mass at the habitable zone ($\sim
1$ AU) is considerably smaller than that of the Earth. 
Nevertheless, terrestrial planets attain most of their 
asymptotic masses, after the gas depletion, through giant 
collisions and merger events  as their velocity dispersion 
increases until they bypass their dynamical isolation.  This 
conjecture is supported by the late Earth's formation epoch 
($\sim 30-60$ Myr as inferred from radioactive isotopes
\citep[e.g.,][]{Yin02,Kleine02}) 
which appears to persist well after the disk gas 
is depleted (on a few Myr time scale).  

In Papers I-IV, this post-oligarchic growth has been taken 
into account with a simple prescription.  In that approach, 
the width of embryos' feeding zone ($\sim 2 e_{\rm esc} a$) 
is assumed to expand with their eccentricity and embryos' 
asymptotic mass is estimated to be (Paper I)
\begin{equation}
M_{\rm e,iso} \simeq
0.68 \eta_{\rm ice}^{3/2} f_{\rm d}^{3/2}
\left(\frac{a}{1{\rm AU}}\right)^{3/2}
\left(\frac{\rho_{\rm d}}{3{\rm gcm}^{-3}}\right)^{1/4} M_{\oplus}.
\label{eq:m_e_iso}
\end{equation}
This approximation naturally reproduces the dependences of the 
outcome on the magnitude of $f_{d,0}$ and $a$. In comparison
the results simulated with our new prescription, the above 
expression for $M_{\rm e, iso}$ slightly underestimates it 
(by a factor of $\sim 2$ or so). 

Figure 4 in Paper V shows that 
a fraction of solar type stars harboring habitable planets
with $M_p \simeq 0.3-10M_{\oplus}$ and $a \simeq 0.75-1.8$AU, 
$\eta_{\oplus} \sim 20\%$ for
$C_1 \la 0.03$ and $\sim 10\%$ for $C_1 \la 0.3$. These values 
do not significantly change by using the present new model.
Thus, with our new prescription, we verify that the 
magnitude of $\eta_\oplus$ is adequately 
reproduced by the simulations presented in Paper V, using 
Eq.~(\ref{eq:m_e_iso}).

\section{Summary and Discussions}
\label{sec:discussions}

In this paper, we introduce a new prescription for our
population synthesis models.  In this revision, we
incorporate, for the first time, a semi analytic prescription
for close scatterings and giant impacts among solid 
planetary embryos (for detailed prescriptions, see Appendix).
Our analytical model quantitatively reproduces statistics
of the asymptotic distributions of mass, semimajor axis and orbital 
eccentricities of bone fide planets
obtained by N-body simulations \citep{Kokubo06},
as shown in \S2.7.

We also take into account the effects of resonant trapping 
by embryos during their migration and when they are stalled 
by migration barriers (\S2.4, 2.5).  With these tools, we are
able to simulate the statistical distributions of solid 
planets' asymptotic mass, semimajor axis and orbital 
eccentricities, in systems without any giant planets.

These simulations are particularly relevant to the 
recently discovered population of close-in super-Earths.
Our results indicate that the emergence of super-Earths
in the proximity of their host stars proceeds through
three stages:
\begin{itemize}
\item In a gas rich environment, embryos accrete planetesimals
in their feeding zones and become dynamically isolated 
(runway/oligarchic growth).
\item Embryos undergo orbital decay and accumulate
close to their host stars 
(migration/stall).
\item After disk gas depletion, embryos perturb and excite 
eccentricities of each other. Eventually, their orbits cross 
and they grow significantly 
beyond the isolation mass through collisions among them
(post-oligarchic giant impacts).
\end{itemize}

During the second and third stages, 
sub-Earth embryos evolve into systems of 
non-resonant, multiple super-Earth planets
through the following mechanisms:
\begin{itemize}
\item The coupled effects of type I migration of embryos, 
termination of the migration near the inner disk edge, 
and resonant trapping between embryos produce a resonantly 
trapped convoy of embryos that extends beyond 0.1AU for the
the cavity inner boundary condition (\S3.2).
\item In the presence of the disk gas, because their orbits 
are stabilized by efficient tidal $e$-damping, embryos' 
individual masses are similar to the critical mass to
resist against type I migration that is $\la M_{\oplus}$.
They do not go into runaway gas accretion phase.
\item Through close scattering and giant impacts after disk 
gas depletion, the resonantly clustered embryos form 
multiple Earths or super-Earths in non-resonant orbits 
around $\sim 0.1$AU in the disks with masses comparable 
to or a few times larger than the MMSN ($f_{d,0} \simeq 1$--3).
\item Merger products attain non-resonant orbits and their 
orbit crossing timescale is longer than the typical age of their 
host stars ($\sim 10$ Gyr). Consequently, their asymptotic dynamical 
architecture is very stable.
\item Although the Earths/super-Earths suffer close scattering, 
their eccentricities are as low as 0.01--0.1 due to efficient 
collision damping and large local Keplerian velocity.
\item Although the asymptotic mass of some super-Earths is larger
than that required to initiate efficient gas accretion, they do
not evolve into gas giants because their final assemblage occur 
only after the residual disk gas is depleted.
\end{itemize}

In disks with $f_{d,0} \la 0.3$, the initial isolation mass 
of embryos interior to the ice line is considerably less than 
that of the Earth.  Type I migration of sub-Earth embryos is
less efficient and the amount of embryos accumulation in the
stellar proximity is limited.  Consequently, close-in 
planets, if formed at all, tend to have masses less than 
that of the Earth.  

For a range of disk accretion rate (comparable to that observed
in typical T Tauri stars), migration of embryos formed in the 
outer regions of protostellar disks is stalled by a barrier 
near the ice line \citep[][Paper V]{Kretke07}.  Although 
the isolation mass at large distances from their host stars is 
substantially larger than that of the Earth, embryos' growth time 
scale is also very long. Embryos may attain masses (mostly made of 
icy material) comparable to that of Uranus and Neptune but they 
would not be able to completely clear their feeding zone (even on 
a time scale of a few Gyr).  Consequently, their eccentricity is 
effectively damped by dynamical friction from the residual 
planetesimals even after the disk gas is depleted.  

In our simulations, we found a relatively large fraction of stars 
bear ice giants, albeit many of these stars do not contain any 
gas giants. While ice giants tend not to cross each other's orbits, 
they, nonetheless, scatter residual planetesimals to large distances 
(beyond 10-100 AU's, analogous to the Kuiper Belt) and continuously 
supply parent bodies of debris disks which emit persistent (over Gyr 
time scale) mid and far IR reprocessed radiation (equivalent to 
cold zodiacal light). In the region interior to the ice line, embryos' 
growth is suppressed by dynamical isolation rather than collisional 
time scales and a few super-Earth oligarchics emerge 
while residual planetesimals between them are mostly acquired by them. 
In this region, post-oligarchic collisions occur stochastically and 
produce dusty fragments which provide sources for the NIR reprocessed
radiation \citep{Kenyon_Bromley04}.  
The occurrence of inner debris disks 
only weakly depends on the host stars' metallicity and mass through 
the asymptotic mass and spacing of the embryos.  

The cases presented here are for disks with modest mass. In 
relatively massive ($f_{d,0} \la 2-3$) or metal-rich disks, 
one or more gas giants may form (Paper I, IV). 
We have not considered here systems which contain gas giants.  The
effects of their secular and resonant perturbation will be considered
in the next paper. Here we draw to the attention that our Solar 
System does not have any planetary body inside Mercury's orbit at 0.4AU.
Yet it must have formed from a disk with $f_{d,0}$ comparable to or larger
than unity ({\it i.e.,} at least in a MMSN).  Many extrasolar planetary 
systems also do not display any sign of close-in super-Earths.
We suggest this dichotomy is due to a dispersion in the sizes of an 
inner disk cavity. The extent of this disk region depends on the 
diverse disk accretion rate and strength of host stars' magnetic field 
during their T Tauri phase of evolution.  

In \S2.5, we suggested that the observed bimodal distribution of spin 
periods of young stars with peaks at about one week and one day 
respectively may be produced by the different strength of the 
coupling \citep{Herbst05}.
If the coupling is stronger than some 
critical value (around stars with a few kilogauss fields), the stellar
spin angular momentum would be transfered to the disk through the 
stellar dipole magnetic field and the ionized disk gas would be 
accreted through the channel flow along the magnetic field lines 
\citep[e.g.,][]{Shu94}. Inside 
the radius ($\sim 0.06-0.1$ AU) where disk gas and stellar spin 
co-rotate (with period of a week), the disk would lose angular momentum 
and be truncated with a magnetospheric cavity \citep[e.g.,][]{Shu94}.

This magnetospheric cavity may not exist or be confined to much smaller 
radii around stars with relatively weak magnetic fields. Inefficiency
of angular momentum transfer would result in much faster stellar spin
(with periods of a day).  In this limit, migrating embryos would be halted
(if at all) so close to their host stars that the star-planet tidal
interaction would lead to further orbital decay and disruption.  We 
simulated such a possibility with models in which no migration barrier 
is imposed at the disk inner boundary (see \S 3.2 and 3.4).  We find
that regions interior to $\sim 0.3$ are effectively cleared by 
early generation of migrating embryos.  

First-generation embryos may have formed in the primordial solar
nebula when the infant Sun had a relatively weak field.  
Extrasolar planetary 
systems without close-in super-Earths may also have formed around
weakly magnetized T Tauri stars. The strength of magnetic coupling may
also evolve with the accretion rate in the disk \citep{Kretke10}.
In a subsequent paper, we will address the magnetic 
coupling process that would play
a key role in the origin of diversity of close-in
super-Earths or Earths.

\vspace{1em} 
\noindent ACKNOWLEDGMENTS.  
We thank Micheal Mayor and Geoff Marcy for useful discussions.
This work is supported by 
JSPS (20244013),
NSFC(10233020), NCET (04-0468), NASA (NNX07A-L13G,
NNX07AI88G, NNX08AM84G), JPL (1270927), and NSF(AST-0908807).

\vspace{1em}
\noindent CORRESPONDENCE should be addressed to S. I. (ida@geo.titech.ac.jp).

\clearpage

\section*{Appendix. Prescription for eccentricity excitation and
merging of embryos in post-oligarchic stage}

We describe here prescriptions for embryos' eccentricity excitations 
and merging process.  Because giant impacts proceed after the depletion 
of disk gas and planetesimals, we consider only mutual gravitational
interactions among the embryos and neglect damping forces 
from disk gas and dynamical friction from planetesimals.

The order of our procedures is:

\noindent
1) set up initial conditions, 

\noindent
2) evaluate, for all pairs of embryos, the crossing time scale 
($\tau_{\rm cross}$) over which sufficiently large eccentricities are 
excited (by their mutual secular perturbations) so that their orbits 
would cross each other,

\noindent
3) evaluate the amount of eccentricity excitation ($e$) and 
semimajor axis change ($\delta a$) of embryos pairs which are 
destined to cross each other's orbits next,

\noindent
4) determine whether these changes may lead to the participating
embryos to undergo secondary orbit crossings with any other neighboring 
embryos,

\noindent
5) compute changes in the $e$ and $a$ for embryos which undergo
secondary orbit crossings,

\noindent
6) make adjustment in $a$ for all embryos participated in orbit crossing
so that their total orbital energy is conserved,

\noindent
7) identify a pair of orbit-crossing embryos which are most likely to 
physically collide with each other,

\noindent
8) create a merged embryo, and

\noindent
9) repeat step 2, until $\tau_{\rm cross}$ of embryo pairs exceed 
the system age.

The detailed prescriptions for each step are as follows:
\begin{enumerate}
\item
The initial mass ($m_j$) and semimajor axis ($a_j$) of $j$th embryo are
determined by oligarchic growth model (see section 2.6).
The eccentricity ($e_j$) of dynamically isolated embryos are assigned
randomly by a Rayleigh distribution with mean values of their Hill
eccentricities (defined as $(m_j/3 M_{\ast})^{1/3}$
where $M_{\ast}$ is the host star mass).
Embryos' initial eccentricities are of the order of 0.001-0.01.
Since eccentricities are significantly excited by embryos' mutual 
interaction after the onset of orbit crossing,
the initial small values of $e_j$ do not affect the results.

\item 
Given $m_j, a_j$, and $e_j$, the crossing time (timescales
on which orbital crossing starts)
of each pair ($i,j$) of the embryos are calculated,
following the fitting formula obtained by \citet{Zhou07} with
some slight modifications:
\begin{equation}
\log \left(\frac{\tau_{\rm cross}}{T_{\rm K}} \right)
= A + B \log \left(\frac{b}{2.3 r_{\rm H}}\right),
\label{eq:tau_cross}
\end{equation}
where $T_{\rm K}$ is Keplerian time at the mean $a (= \sqrt{a_i a_j})$
of the pair,
$b = | a_i - a_j |$, $r_{\rm H} = ((m_i+m_j)/3 M_{\ast})^{1/3}a$, and
\begin{equation}
\begin{array}{l}
A = -2 + e_0 - 0.27 \log \mu, \\
B = 18.7 + 1.1 \log \mu - (16.8 + 1.2 \log \mu)e_0, \\
{\displaystyle e_0 = \frac{1}{2} \frac{(e_i + e_j)a}{b}}, \\
{\displaystyle \mu = \frac{(m_i + m_j)/2}{M_{\ast}}}.
\end{array}
\end{equation}

\item
The pair with the shortest orbit crossing time ($\tau^*_{\rm cross}$) is 
assumed to undergo close encounters before any other pairs.  During these
close encounters, embryos' excited velocity dispersion is limited
by surface escape velocity of the perturber $v_{{\rm esc},j}$
\citep{Safronov69,Palmer}.
The corresponding eccentricity is
\begin{equation}
e_{{\rm esc},ij} = \frac{v_{{\rm esc},ij}}{v_{\rm K}}
= \frac{\sqrt{2G(m_i+m_j)/(R_i + R_j)}}{\sqrt{G M_{\ast}/a}}
\simeq 0.28 \left(\frac{m_i+m_j}{M_{\oplus}}\right)^{1/3}
  \left(\frac{\rho}{3{\rm gcm}^{-3}}\right)^{1/6}
  \left(\frac{a}{1{\rm AU}}\right)^{1/2}, 
\label{eq:e_esc}
\end{equation}
where $v_{\rm K}$ is Keplerian velocity,
and $R_j$ is the physical radius calculated from $m_j$
with bulk density $\rho$ of 3 gcm$^{-3}$ inside the ice line 
and 1 gcm$^{-3}$ outside it.
The eccentricity change is partitioned according to
the mass of interacting bodies
\citep{Nakazawa_Ida88, Nakazawa_Ida89} and 
the eccentricity distribution is relaxed to 
a Rayleigh distribution \citep{Ida_Makino92}.
Thus, the individual excited eccentricities are given by
\begin{equation}
e_j = \frac{m_i}{m_i + m_j} e_{{\rm esc},ij} {\cal R},
\label{eq:close_scat}
\end{equation}
where ${\cal R}$ is a random number produced by a Rayleigh 
distribution with the mean value of unity. Changes in the semimajor 
axis associated with the eccentricity excitation are assumed to 
be $\pm e_j a_j$. Sign of the change is chosen to ensure that
orbital separation between the pair is widened after their orbit
crossing.

\item 
Energy and angular momentum exchanges during the close encounters
between any pair of orbit-crossing embryos may induce them to cross 
the orbits of other nearby embryos. We classify all the affected 
embryos by their overlapping (relative to the modified pair) radial 
excursion (epicycle amplitude) into closely interacting groups.  
These groups generally contain several ($\ge 3$) embryos.
We assume that the most massive embryo in each group dominates
the outcome of successive close encounters within the group.  
The characteristic post-encounter eccentricity ($e_j$) for 
all but the most-massive embryos is given by equation 
(\ref{eq:close_scat}) in which $i$ is used to represent the most 
massive embryos in the group.  We also assume the post-encounter
eccentricity of the most-massive embryo is determined mostly
by the second-most-massive embryo. 

In general, successive close encounters lead to radial diffusion 
of the orbit-crossing embryos.  In our prescription, 
changes in embryos' semimajor axis are given by values chosen
$W_j e_j a_j$.  The statistical weight $W_j$ (which has a skewed
distribution within a limited range between -1 and 1) is introduced
to reflects the kinematic distribution of the orbit crossing embryos.  
If the total mass ($M_{\rm in}$) of all the encountering embryos in 
the group inside a particular embryo's orbit is more than that
($M_{\rm out}$)  outside it, its probability of orbital expansion 
would be larger than that of orbital decay. In order to simulate
this effect, we introduce a parameter $f_j = C_j (M_{\rm in} -
M_{\rm out})/(M_{\rm in} +M_{\rm out})$ and set $W_j = f_j +
(1 - | f_j |) * R_j$ where $R_j$ is a random number with a uniform
distribution in the range of $[-1, 1]$. With the results of N-body
simulations, we also calibrate the magnitude of a constant $C_j 
\sim 2/3$. 

Finally, close-encounters also excite embryos' inclination.  
Magnitude of the orbital inclinations in radian may be about 
half of that of eccentricities 
\citep{Ida_Makino92}. However, the orbital inclinations do not 
directly affect the condition for orbit crossing, we are not 
concerned with embryos' inclinations in this paper.

\item
After the relative changes in the semimajor axes of all the 
orbit crossing embryos have been evaluated, we renormalize 
their semimajor axes by a constant numerical factor so that the
total orbital energy of all the embryos in the group is conserved.
In general, the modification of the semimajor axes introduced by
this renormalization procedure is less than a percent order.
This simple algorithm does introduce artificial changes in the 
total angular momentum of the encountering embryos.  

Although, with a more complicated non-uniform renormalization
factor, we can also conserve the total angular momentum of 
all the embryos in the group, we opted to limit the degrees of
freedom at the expense of a small change in the total angular 
momentum of the system.  After multiple close encounters and 
merger events, the total departure from angular momentum 
conservation of the entire system is limited to within 10\% 
in typical models.

\item
Close encounters also leads to physical collisions.  We 
assume all physical collisions are cohesive and lead to 
merger events. Among the embryos in the orbit crossing group, 
a colliding pair ($k,l$) is chosen.  However, after the 
impulsive momentum changes during repeated close-encounters
not every chosen pairs from this group would retain overlapping 
orbits. If the radial excursions of any chosen pair do not 
overlap, another pair would be chosen until all available 
candidates are exhausted. (For example, after a close encounter 
between embryos 1 and 2, the former may cross the orbits of 
embryos 3 and 4 whereas the latter may cross the orbits of 
embryos 3 and 5.  If collisional candidates are chosen to be
embryos 1 and 5, this pair would be rejected.)

In order to take into account various factors which contributes 
to the collisional probabilities, we select the appropriate embryos 
with a statistically weighted random number.  N-body simulations 
show that collision frequency decreases with embryos' semimajor 
axes (because Keplerian period is longer and embryos' spatial 
density is lower at larger semimajor axis).  We adopt a weighted 
collisional probability which is proportional to $a^{-3}$.

\item 
A merged embryo is created from each colliding pair. Since
we neglect fragmentation and rebound, the merged embryo
acquires the total mass of the colliding pair.  In the direct 
N-body simulations, the center of inertia and the total momentum 
are conserved after each cohesive collision.  In our prescription,
we do not follow the exact location and kinematics of the merger
product. Instead, we use conservation laws to determine orbital
elements of the merged embryos.

In principle, a small fraction of orbital angular momentum
is transformed into spin angular momentum after each collision.
However, the spin angular momentum is usually negligible.
Total kinetic energy of the colliding embryos is also not 
conserved, because the collision dissipates kinetic energy of 
the relative motion and the merging changes binding energy.
However, if their velocity dispersions are smaller than
their two-body surface escape velocity ($v_{{\rm esc},kl}$), 
the collisional dissipation energy is 
approximately equal to the change in the binding energy.
In practice, the total orbital energy of the colliding pair is 
approximately conserved during the collision as well as
the total angular momentum.

Assuming the conservation of orbital energy, 
the semimajor axis of the merged body ($a_{kl}$) is given by
\begin{equation}
\frac{m_k + m_l}{a_{kl}} = \frac{m_k}{a_{k}} + \frac{m_l}{a_{l}}. 
\end{equation}
In the case of $e \ll 1$, it is difficult to accurately evaluate the 
eccentricity of the merged body
from the angular momentum conservation. Here, we assume 
that after each collision, the mass-weighted total Laplace-Runge-Lenz 
vector is conserved. This assumption holds for collisions between
embryos with nearly circular Keplerian motion ({\it i.e.} $e \ll 1$)
\citep{Nakazawa_Ida88,Nakazawa_Ida89}.  In this limit, 
the eccentricity of the merged body ($e_{kl}$) is given by
\begin{equation}
\begin{array}{l}
(m_k + m_l) e_{kl} \cos \varpi_{kl} = m_k e_{k} \cos \varpi_{k} +
m_l e_{l} \cos \varpi_{l}, \\
(m_k + m_l) e_{kl} \sin \varpi_{kl} = m_k e_{k} \sin \varpi_{k} +
m_l e_{l} \sin \varpi_{l}, 
\end{array}
\label{eq:LRL_sum}
\end{equation}
where $\varpi$'s are longitudes of periastron at the collision.

Due to the secular perturbation between orbit-crossing embryos,
$\varpi_{k}$ and $\varpi_{l}$ generally precess independently
over all angles. In principle, both longitudes can be chosen 
with uniformly distributed random phase (between 0 and $2 \pi$).
However, even with overlapping regions of radial excursion, 
the orbits of a pair of eccentric embryos would only  
cross each other if the relative angle between their
longitudes of periastron ($\theta = \varpi_k - \varpi_l$) is within
some limited range. For example, eccentric embryos would not 
cross each other's orbits if their longitudes of periastron are 
aligned with $\theta \simeq 0$.  In contrast, orbit crossing 
generally occurs between embryos with $\theta$ in some limited 
range centered at 180 degrees, provided $e_k a_k$ and $e_j a_j$ 
are larger than $| a_k - a_l | $.  

In our prescription, we randomly generate a value of  
$\varpi_l$ between 0 and $2 \pi$.  We assume secular 
perturbation generally induces the longitudes of periastron 
of any pairs of embryos with overlapping orbits to precess
and circulate so that their $\theta$ can always enter into 
the range ($\Delta \theta$) which allows them to collide.  
We first compute the magnitude
of $\Delta \theta$ with independent Keplerian orbits for 
the neighboring embryos.  We then take into account that 
embryos' gravitational perturbation on each other tends to 
slightly broaden this range. The value of $\theta$ at the 
collision is randomly chosen from a value within the range 
$\Delta \theta$. The value of $\varpi_k$ is then specified 
by $\varpi_l$ and $\theta$.  Since the preferred value for 
$\theta$ is $\sim \pi$, 
the summation in the r.~h.~s. of eq.~(\ref{eq:LRL_sum})
usually results in cancellation, so that
$e_{kl}$ is often significantly smaller than $e_k$ and $e_l$. 


\item 
For updating the system time, we need to take into consideration
1) the time interval for neighboring embryos to cross each other's 
orbits ($\tau_{\rm cross} = \tau^*_{\rm cross}$ in equation 
\ref{eq:tau_cross}) and that required for embryos to collide with 
each other $\tau_{\rm collide}$.  Before each collision, embryos 
with overlapping orbits undergo repeated close scattering as $\theta$ 
circulates in and out the range of $\Delta \theta$. The magnitude of 
$\tau_{\rm collide}$ for embryos with overlapping orbits is determined 
by their area filling factor (see Paper I).  N-body simulations indicate 
that at a fraction of AU, $\tau_{\rm collide} \sim 10^{5.5} - 10^{6.5} 
T_{\rm K}$ for Earth-size embryos.  

For pairs with moderately large initial 
separation ($b \ga 5 r_{\rm H}$), 
$\tau_{\rm cross} \gg \tau_{\rm collide}$ and it is adequate to 
update the evolution of the system from $t$ to $t + \tau^*_{\rm cross}$.
Pairs with small initial separation ($b \la 5 r_{\rm H}$) are closely
packed with wide range of $\Delta \theta$. Since it takes longer for them
to collide than to cross each other's orbit, we update the evolution with
a randomly generated $\tau_{\rm collide}$ which has a uniform 
distribution in the range $(10^{5.5}-10^{6.5}) T_{\rm K}$.
Thus, the new system time is given by 
$t + \tau^*_{\rm cross}$ for $\tau^*_{\rm cross} > 10^{5.5} T_{\rm K}$
and $t + \tau^*_{\rm collide}$ otherwise.

After some collisional events, it is possible for the merger product
to become dynamically isolated from all other embryos in the group. 
N-body simulations also indicate that following other cohesive collisions, 
several additional merger events may follow in rapid succession. 
For example, in a group of 4 orbit-crossing embryos, embryos1 and 2 may merge first.  The newly form embryo may attain a new orbit 
which continues to overlap those of embryos 3 and 4.  In that case,
it is possible for the newly merged embryo to subsequently coagulate 
with embryos 3 or 4 or both.  In order to take these possibilities 
into account, we repeat step 2 to 7 for all group members after each 
merger event, until their $\tau^*_{\rm cross}$ (including the collisional
lag time) has exceeded the integration time.
  
\end{enumerate}

In step 7, we indicate that the merger product attains an
eccentricity ($e_{kl}$) which is often significantly smaller 
either than $e_k$ and $e_l$.  This apparent eccentricity
damping is due to the effect that merger events can only occur
for a limited range between the longitude of periastron of 
the colliding embryos. Consequently, embryos' asymptotic 
eccentricities are usually smaller than $e_{\rm esc}$. 
This inference is consistent with
the results of previous N-body simulations \citep{Kokubo06}.
  
In the prescription presented here, the only free empirical
parameters are i) the weighted changes in the semimajor axis
in step 4 and ii) the broadened range of $\Delta \theta$
in step 6.  All the other procedures are based on fundamental 
dynamics although some of their efficiencies are approximated.   
The weighted $a$ changes and the broadening of the phase range
are also qualitatively based on celestial mechanics.  The quantitative 
parameters for these processes are determined by comparison with 
the results obtained by N-body simulations \citep{Kokubo06}.  Note that 
the same parameters are applied for all models in the present paper and
we do not individually carry out fine tuning for different disk conditions.

{}

\clearpage

\clearpage

\clearpage

\clearpage

\clearpage

\clearpage

\clearpage

\clearpage

\clearpage

\clearpage

\begin{thebibliography}{}

\bibitem[Aarseth et al.(1993)]{Palmer}
Aarseth, S.~J., Lin, D.~N.~C., Palmer, P.~L. 1993, ApJ, 403, 351

\bibitem[Artymowicz(1993)]{Artymowicz93}
Artymowicz, P. 1993, \apj, 419, 166

\bibitem[Baruteau \& Lin(2010)]{Baruteau10}
Baruteau, C. \& Lin, D. N. C. 2010,
\apj, 709, 759

\bibitem[Beckwith \& Sargent(1996)]{Beckwith96}
Beckwith, S.~V.~W. \& Sargent, A.~I., 1996,
Nature, 383, 139

\bibitem[Bodenheimer \& Pollack(1986)]{Bodenheimer86}
Bodenheimer, P., \& Pollack, J. B. 1986, Icarus, 67, 391

\bibitem[Bodenheimer et al.(2000)]{Bodenheimer00}
Bodenheimer, P., Hubickyj, O. \& Lissauer, J. J. 2000,
Icarus, 143, 2

\bibitem[Bryden et al.(2009)]{Bryden09}
Bryden, G. et al. 2009,
\apj, 705, 1226

\bibitem[Chambers et al.(1996)]{Chambers96}
Chambers, J. E., Wetherill, G. W., Boss, A. P. 1996, Icarus, 119, 261

\bibitem[Chiang \& Goldreich(1997)]{Chiang97}
Chiang, E. I. \& Goldreich, P. 1997,
\apj, 490, 368

\bibitem[Cieza \& Baliber (2007)]{CB07}
Cieza, L. \& Baliber,N. 2007, \apj, 671, 605

\bibitem[Cumming et al.(2008)]{Cumming08}
Cumming, A., Butler, R. P., Marcy, G. W., Vogt, S. S.,
Wright, J. T., Fischer, D. 2008, PASP, 120, 531

\bibitem[Duncan \& Levison (1997)]{Duncan97}
Duncan, M. J. \& Levison, H. F. 1997,
Science, 276, 1670

\bibitem[Fischer \& Valenti (2005)]{Fischer05}
Fischer, D. A. \& Valenti, J. A. 2005.
\apj, 622, 1102

\bibitem[Garaud \& Lin(2007)]{Garaud07}
Garaud, P. \& Lin, D. N. C. 2007,
\apj, 654, 606

\bibitem[Goldreich \& Lynden-Bell(1969)]{Goldreich69}
Goldreich, P. \& Lynden-Bell, D. 1969, \apj, 156, 59

\bibitem[Goldreich \& Tremaine(1982)]{GT82}
Goldreich, P., \& Tremaine, S. 1982, ARA\&A, 20, 249

\bibitem[Goldreich et al.(2004)]{Goldreich04}
Goldreich, P., Lithwick, Y. \& Sari, R. 2004, ARA\&A, 42, 549

\bibitem[Greaves et al.(2007)]{Greaves}
Greaves, J. S., Fischer, D. A., Wyatt, M. C., Beichman, C. A.,
\& Bryden, G. 2007, MNRAS, 378, L1

\bibitem[Haisch et al.(2001)]{Haisch}
Haisch, K. E., Lada, E. A. \& Lada, C. J. 2001, ApJ, 553, L153

\bibitem[Hasegawa \& Nakazawa(1990)]{Hasegawa90}
Hasegawa, M. \& Nakazawa, K. 1990, A\&A, 227, 619

\bibitem[Hayashi(1981)]{Hayashi81}
Hayashi, C. 1981, Prog. Theor. Phys. Suppl., 70, 35

\bibitem[Herbst \& Mundt(2005)]{Herbst05}
Herbst, W. \& Mundt, R.
\apj, 633, 967

\bibitem[Ida \& Makino(1992)]{Ida_Makino92}
Ida, S. \& Makino, J. 1992. Icarus 96, 107

\bibitem[Ida \& Lin(2004a)]{IL04a}
Ida, S. \& Lin, D. N. C. 2004,
\apj, 604, 388 (Paper I)

\bibitem[Ida \& Lin(2004b)]{IL04b}
Ida, S. \& Lin, D. N. C. 2004,
\apj, 616, 567 (Paper II)

\bibitem[Ida \& Lin(2005)]{IL05}
Ida, S. \& Lin, D. N. C. 2005,
\apj, 626, 1045 (Paper III)

\bibitem[Ida \& Lin(2008a)]{IL08a}
Ida, S. \& Lin, D. N. C. 2008a,
\apj, 673, 487 (Paper IV)

\bibitem[Ida \& Lin(2008b)]{IL08b}
Ida, S. \& Lin, D. N. C. 2008b,
\apj, 685, 584 (Paper V)

\bibitem[Ikoma et al.(2000)]{Ikoma00}
Ikoma, M., Nakazawa, K. \& Emori, E. 2000, 
\apj, 537, 1013

\bibitem[Ikoma et al.(2001)]{Ikoma01}
Ikoma, M., Emori, E. \& Nakazawa, K. 2001,
\apj, 553, 999

\bibitem[Iwasaki et al.(2002)]{Iwasaki02}
Iwasaki, K., Emori, H., Nakazawa, K. \& Tanaka, H. 2002,
PASJ, 54, 471

\bibitem[Kato et al.(2009)]{Kato09}
Kato, M. T., Nakamura, K., Tandokoro, R., Fujimoto, M.
\& Ida, S. 2009, \apj, 691, 1697

\bibitem[Kennedy et al.(2006)]{Kennedy06}
Kennedy, G. M., Kenyon, S. J. \& Bromley, B.C. 2006, \apjl, 650, L139

\bibitem[Kennedy \& Kenyon(2008)]{Kennedy08}
Kennedy, G. M. \& Kenyon, S. J. 2008, \apj, 682, 1264

\bibitem[Kenyon \& Bromley(2004)]{Kenyon_Bromley04}
Kenyon, S.J. \& Bromley, B.C. 2004, \apj, 602, 133

\bibitem[Kleine et al.(2002)]{Kleine02}
Kleine, T., Munker, C., Mezger, K., Palme, H., 2002,
Nature, 418, 952

\bibitem[Kokubo \& Ida(1996)]{KI96}
Kokubo, E. \& Ida, S. 1996, Icarus, 123, 180

\bibitem[Kokubo \& Ida(1998)]{KI98}
---------. 1998, Icarus, 131, 171

\bibitem[Kokubo \& Ida(2002)]{KI02}
---------. 2002, \apj, 581, 666

\bibitem[Kokubo et al.(2006)]{Kokubo06}
Kokubo, E., Kominami, J. \& Ida, S. 2006, \apj, 642, 1131

\bibitem[Koller et al.(2003)]{Koller03}
Koller, J., Li, H. \& Lin, D. N. C. 2003,
\apjl, 596, L91

\bibitem[Kominami \& Ida(2002)]{Kominami02}
Kominami, J. \& Ida, S. 2002, Icarus, 157, 43

\bibitem[Konigl(1991)]{Konigl91}
Konigl, A. 1991, \apjl, 370, L39

\bibitem[Kretke \& Lin(2007)]{Kretke07}
Kretke, K. A. \& Lin, D. N. C. 2007,
ApJ, 664, L55

\bibitem[Kretke et al.(2009)]{Kretke09}
Kretke, K. A., Lin, D. N. C.,
Garaud, P. \& Turner, N. J. 2009,
ApJ, 690, 407

\bibitem[Kretke \& Lin(2010)]{Kretke10}
Kretke, K. A. \& Lin, D. N. C. 2010,
in preparation

\bibitem[Laine et al.(2008)]{Laine08}
Laine, R. O., Lin, D. N. C. \& Dong, S. 2008.
\apj 685, 521

\bibitem[Laine \& Lin(2010)]{Laine10}
Laine, R. O. \& Lin, D. N. C. 2010, in preparation

\bibitem[Laughlin et al.(2004)]{Laughlin04}
Laughlin, G., Steinacker, A. \& Adams, F.C. 2004.
\apj 608, 489

\bibitem[Lecar et al.(2006)]{Lecar06}
Lecar, M., Podolak, M., Sasselov, D. \& Chiang, E. 2006,
\apj, 640, 1115

\bibitem[Li et al.(2009)]{Li09}
Li, H., Lubow, S. H., Li, S. \& Lin, D. N. C. 2009,
\apj, 690, 52

\bibitem[Lin et al.(1996)]{Lin96}
Lin, D.~N.~C., Bodenheimer, P. \& Richardson, D. 1996,
Nature, 380, 606

\bibitem[Lin \& Papaloizou(1985)]{LP85}
Lin, D.~N.~C. \& Papaloizou, J.~C.~B. 1985,
Protostars and Planets II, 
ed. D.~C. Black \& M.~S.Matthew (Tucson: Univ. of Arizona Press), 981

\bibitem[Lin \& Papaloizou(1993)]{LP93}
Lin, D. N. C., \& Papaloizou, J. C. B. 1993, 
in Protostars and
Planets III, ed. E. H. Levy and J. I. Lunine (Tucson:Univ. of Arizona
Press), 749

\bibitem[Lissauer(1987)]{Lissauer87}
Lissauer, J. 1987.
Icarus, 69, 249

\bibitem[Lynden-Bell \& Pringle(1974)]{Lynden-Bell74}
Lynden-Bell, D. \& Pringle, J. E. 1974, MNRAS, 168, 603

\bibitem[Mardling \& Lin(2004)]{Mardling_Lin2004}
Mardling, R.A. \& Lin, D.N.C. 2004, \apj, 614, 955

\bibitem[Masset et al.(2006)]{Masset06}
Masset, F. S., Morbidelli, A., Crida, A. \& Ferreira, J.
2006, \apj, 703, 857

\bibitem[Masset \& Casoli(2009)]{Masset09}
Masset, F. S. \& Casoli, J. 2009,
\apj, 703, 857

\bibitem[Mayor {\it et.al}(2009)]{Mayor09}
Mayor, M. {\it et al.} 2009, A\&A, 507, 487

\bibitem[McNeil et al.(2005)]{McNeil}
McNeil, D., Duncan, M. \& Levison, H. 2005. 
\aj, 130, 2884

\bibitem[Marzari \& Weidenschilling (2002)]{MW02}
Marzari, F. \& Weidenschilling, S. 2002, Icarus, 156, 570

\bibitem[Mizuno(1980)]{Mizuno80}
Mizuno, H. 1980.
Prog. Theor. Phys. 

\bibitem[Mordasini et al.(2009)]{Mordasini09}
Mordasini, C., Alibert, Y. \& Benz, W. 2009.
A\&A, 501, 1161

\bibitem[Murray \& Dermott(2000)]{Murray_Dermott}
Murray, C. D. \& Dermott, S. F. 2000,
Solar System Dynamics, Cambridge University Press

\bibitem[Nagasawa et al.(2005)]{Nagasawaetal05}
Nagasawa, M., Lin, D.N.C., \& Thommes, E. 2005, \apj 635, 578

\bibitem[Nakazawa \& Ida(1988)]{Nakazawa_Ida88}
Nakazawa, K. \& Ida, S. 1988. 
Prog. Theor. Phys. Suppl. 96, 167

\bibitem[Nakazawa \& Ida(1989)]{Nakazawa_Ida89}
Nakazawa, K. \& Ida, S. 1989. 
A\&A, 220, 293

\bibitem[Nelson(2005)]{Nelson05}
Nelson, R.P. 2005. A\&A, 443, 1067

\bibitem[O'Brien et al.(2006)]{O'Brien06}
O'Brien, D., Morbidelli, A. \& Levison, H. F. 2006,
Icarus, 184, 39

\bibitem[Ogihara \& Ida(2009)]{Ogihara09}
Ogihara, M. \& Ida, S. 2009. submitted

\bibitem[Ogihara et al.(2010)]{Ogihara10}
Ogihara, M., Duncan, M. \& Ida, S. 2010, submitted

\bibitem[Paardekooper et al.(2010)]{Paardekooper10}
Paardekooper, S.-J., Baruteau, C., Crida, A. \& Kley, W.
2009, MNRAS, 401, 1950

\bibitem[Pan \& Sari(2005)]{Pan05}
Pan, M. \& Sari, R. 2005, Icarus, 173, 342

\bibitem[Pollack et al.(1994)]{Pollack94}
Pollack, J. B., Hollenbach, D., Beckwith, S., Simonelli, D. P.,
Roush, T., \& Fong, W. 1994, \apj, 421, 615

\bibitem[Pollack et al.(1996)]{Pollack96}
Pollack, J. B., Hubickyj, O., Bodenheimer, P., Lissauer, J. J.,
Podolak, M., \& Greenzweig, Y. 1996, Icarus, 124, 62

\bibitem[Rebull et al.(2006)]{reb06}
Rebull, L. M., Stauffer, J. R., Megeath, S. T., Hora, J. L. \& Hartmann, L. 
2006, \apj, 646, 297

\bibitem[Safronov(1969)]{Safronov69}
Safronov, V. 1969,
Evolution of the Protoplanetary Cloud and Formation of
the Earth and Planets (Moscow: Nauka Press)

\bibitem[Shakura \& Sunyaev(1973)]{alpha}
Shakura, N. I. \& Sunyaev, R. A. 1973, A\&A, 24, 337

\bibitem[Schlaufman et al.(2009)]{Schlaufman09}
Schlaufman, K. C., Lin, D. N. C. \& Ida, S. 2009,
\apj, 691, 1322

\bibitem[Schlaufman et al.(2010)]{Schlaufman10}
Schlaufman, K. C., Lin, D. N. C. \& Ida, S. 2010,
submitted

\bibitem[Shu et al.(1994)]{Shu94}
Shu, F., Najita, J., Ostriker, E.,
Wilkin, F., Ruden, S. \& Sunasa, L.
\apj, 429, 781

\bibitem[Stewart \& Ida(2000)]{Stewart00}
Stewart, G. R. \& Ida, S. 2000,
Icarus, 143, 28

\bibitem[Stewart \& Leinhardt(2009)]{Stewart09}
Stewart, S. T. \& Leinhardt, Z. M. 2009,
\apjl, 691, L133

\bibitem[Takeuchi \& Artymowicz(2001)]{Takeuchi01}
Takeuchi, T. \& Artymowicz, P. 2001, \apj, 557, 990

\bibitem[Tanaka et al.(2002)]{Tanaka02}
Tanaka, H., Takeuchi, T. \& Ward, W. 2002, \apj, 565, 1257

\bibitem[Tanaka \& Ward(2004)]{Tanaka04}
Tanaka, H. \& Ward, W. 2004, \apj, 602, 388

\bibitem[Terquem \& Papaloizou(2007)]{Terquem07}
Terquem, C. \& Papaloizou, J. C. B. 2007, 
\apj, 654, 1110

\bibitem[Thommes et al.(2008a)]{Thommes08}
Thommes, E. W., Matsumura, S. \& Rasio, F. A., 2008,
Science, 321, 814

\bibitem[Thommes et al. (2008b)]{Thommesetal08}
Thommes, E., Nagasawa, M., \& Lin, D.N.C., 2008, \apj 676, 728

\bibitem[Ward(1986)]{Udry_Santos}
Udry, S. \& Santos, N. C. 2007, ARA\&A, 45, 397

\bibitem[Ward(1986)]{Ward86}
Ward, W. 1986, Icarus, 67, 164

\bibitem[Ward(1993)]{Ward93}
Ward, W. 1993, Icarus, 106, 274

\bibitem[Wyatt(2008)]{Wyatt08}
Wyatt, M. C. 2008, ARA\&A, 46, 339

\bibitem[Yin et al.(2002)]{Yin02}
Yin, Q. et al. 2002, Nature, 418, 949

\bibitem[Zhou et al.(2007)]{Zhou07}
Zhou, J., Lin, D. N. C. \& Sun, Y. 2007,
\apj, 666, 423

\end{thebibliography}
\end{document}